\documentclass[a4paper,11pt]{article}
\pdfoutput=1

\usepackage{jcappub}
\usepackage[T1]{fontenc}
\usepackage{physics}

\newcommand{\Ito}{It\^{o}}
\newcommand{\tplus}{t_+}
\newcommand{\Nplus}{N_+}
\newcommand{\R}{\mathcal{R}}
\renewcommand*{\vec}[1]{\mathbf{#1}}
\newcommand{\MPl}{M_\mathrm{Pl}}

\title{It\^{o}, Stratonovich, and zoom-in schemes in stochastic inflation}
\author{Eemeli Tomberg}
\affiliation{Consortium for Fundamental Physics, Physics Department,\\Lancaster University, Lancaster LA1 4YB, United Kingdom.}
\emailAdd{e.tomberg@lancaster.ac.uk}

\abstract{
The It\^{o} and Stratonovich approaches are two ways to integrate stochastic differential equations. Detailed knowledge of the origin of the stochastic noise is needed to determine which approach suits a particular problem. I discuss this topic pedagogically in stochastic inflation, where the noise arises from a changing comoving coarse-graining scale or, equivalently, from `zooming in' into inflating space. I introduce a zoom-in scheme where deterministic evolution alternates with instantaneous zoom-in steps. I show that this alternating zoom-in scheme is equivalent to the It\^{o} approach in the Markovian limit, while the Stratonovich approach doesn't have a similar interpretation. In the full non-Markovian setup, the difference vanishes. The framework of zoom-in schemes clarifies the relationship between computations in stochastic inflation, linear perturbation theory, and the classical $\Delta N$ formalism. It informs the numerical implementation of stochastic inflation and is a building block for a first-principles derivation of the stochastic equations.
}

\begin{document}
\maketitle
\flushbottom

\section{Introduction}
\label{sec:intro}
Observations show that the large-scale universe is homogeneous and isotropic, with only small, Gaussian perturbations, typically one part in $10^5$ \cite{Planck:2018jri}. These features are explained by cosmic inflation, an early period of accelerating expansion driven by a scalar field called the inflaton. The expansion smooths out prior inhomogeneities but also stretches quantum vacuum fluctuations to observable scales. Due to their smallness, the typical fluctuations can be described by linear perturbation theory, by adding small perturbations to the metric and energy-momentum tensors on top of the homogeneous Friedmann--Lema\^{i}tre--Robertson--Walker (FLRW) solution and solving the Einstein equation to first order. The observables can be condensed into the \emph{curvature perturbation} and its power spectrum, which freeze to constant values at super-Hubble scales.

The \emph{separate universe approximation} \cite{Salopek:1990jq, Wands:2000dp} provides another way to describe inflationary perturbations. Each super-Hubble patch of inflating space behaves like an independent, local FLRW universe. The super-Hubble perturbations arise as differences between the different patches---in particular, the different amounts of expansion map to curvature perturbations. The mapping is called the $\Delta N$ formalism \cite{Sasaki:1995aw, Sasaki:1998ug, Wands:2000dp, Lyth:2004gb}. Incorporating the short-scale quantum fluctuations gives rise to stochastic inflation \cite{Starobinsky:1986fx}, where the local inflaton evolution is affected by stochastic noise and follows a stochastic differential equation (SDE), resembling Brownian motion. For an introduction to SDEs, see \cite{Lawler:2014}.

When applied to the typical perturbations, the `stochastic $\Delta N$ formalism' reproduces the results of linear perturbation theory \cite{Fujita:2013cna, Fujita:2014tja, Vennin:2015hra}. However, the separate universe approximation is non-perturbative, and the stochastic approach captures super-Hubble evolution beyond linear order, including non-Gaussian features. This is important for the strongest perturbations that collapse into primordial black holes (PBHs) \cite{Carr:1974nx,Carr:1975qj}. Over recent years, stochastic inflation has gained popularity as the tool of choice for studying PBH statistics, see, e.g., \cite{Pattison:2017mbe, Cruces:2018cvq, Ezquiaga:2018gbw, Biagetti:2018pjj, Firouzjahi:2018vet, Ezquiaga:2019ftu, Pattison:2019hef, Prokopec:2019srf, Ando:2020fjm, De:2020hdo, Figueroa:2020jkf, Ballesteros:2020sre, Vennin:2020kng, Cruces:2021iwq, Rigopoulos:2021nhv, Pattison:2021oen, Achucarro:2021pdh, Hooshangi:2021ubn, Tomberg:2021xxv, Figueroa:2021zah, Tada:2021zzj, Cruces:2022imf, Ahmadi:2022lsm, Animali:2022otk, Jackson:2022unc, Tomberg:2022mkt, Tomberg:2023kli, Asadi:2023flu, Briaud:2023eae, Mishra:2023lhe, Raatikainen:2023bzk, Vennin:2024yzl, Jackson:2024aoo}.

An often overlooked aspect of stochastic inflation is the different approaches to stochastic integration. Unlike ordinary differential equations, SDEs can be integrated in many incompatible ways, producing different solutions. The \emph{\Ito} and \emph{Stratonovich} approaches are the most common ones \cite{Ito:1944, Stratonovich:1966, ItoStratoWebpage}, corresponding to a simple Euler integration scheme and a more complicated scheme that preserves the chain rule. In physical systems, these lead to different predictions (in stochastic inflation, slightly different curvature perturbations), and the right approach depends on the details of the physics that causes the stochasticity \cite{Kampen:1981}. For a recent review of the debate between the two approaches in general physical systems, see \cite{Mannella:2012}.

A handful of works have considered the issue in stochastic inflation from the point of view of various symmetry and consistency arguments \cite{Salopek:1990re, Mezhlumian:1991hw, Linde:1993xx, Winitzki:1995pg, Garriga:1997ef, Vilenkin:1999kd, Winitzki:2008zz, Fujita:2014tja, Vennin:2015hra, Tokuda:2017fdh, Pinol:2018euk, Pinol:2020cdp}. However, a physical interpretation of the differences is lacking. In this paper, I point out that the fluctuations in stochastic inflation arise from `zooming in' to ever shorter length scales in comoving coordinates. Different descriptions of this zoom-in process may give rise to different interpretations of the stochastic equation. Since the zooming is independent of the classical (non-stochastic) drift, the most obvious zoom-in scheme consists of alternating independent steps of classical evolution and quantum kicks. I show that, in the usual de Sitter approximation with Markovian noise and in the limit of small time steps, this reduces to the \Ito\ approach; I demonstrate this with a numerical computation of $\phi^2$ inflation.
If the noise is computed beyond the de Sitter approximation by evolving the short-wavelength modes explicitly in the local background, as done in recent numerical studies \cite{Figueroa:2021zah, Figueroa:2020jkf}, its Markovian nature is lost. I show the system can still be recast into the form of a traditional SDE by promoting the mode functions to an equal position with the local background quantities. The \Ito\ and Stratonovich approaches coincide with each other and with the alternating zoom-in scheme in this setup.

In addition to the new results, I discuss stochastic calculus and the different approaches to stochastic integration in a pedagogical way, presenting heuristic derivations of many standard results.

The paper is organized as follows: In Section~\ref{sec:stochastic}, I introduce the stochastic inflation formalism and discuss the zoom-in schemes. In Section~\ref{sec:markovian}, I describe the difference between the \Ito\ and Stratonovich approaches and interpret the alternating zoom-in scheme as an \Ito\ integral in the Markovian slow-roll limit. Section~\ref{sec:non_markovian} goes beyond slow roll and discusses the zoom-in process in the non-Markovian case, and Section~\ref{sec:discussion} is reserved for discussion and comparison to previous studies. Finally, Appendix~\ref{sec:stochastic_formulas} reviews some additional results in stochastic calculus. I work with the metric convention $(-,+,+,+)$ and use natural units where $c=\hbar = \MPl=1$.

\newpage

\section{Zooming into the inflationary perturbations}
\label{sec:stochastic}
Consider single-field inflation driven by the inflaton $\phi$ with a potential $V(\phi)$ and the canonical action
\begin{equation} \label{eq:S}
    S = \int \dd^4 x \sqrt{-g}\qty[\frac{1}{2} R - \frac{1}{2}\partial^\mu\phi\partial_\mu\phi - V(\phi)] \, .
\end{equation}
At the background level, the metric takes the FLRW form
\begin{equation} \label{eq:ds2_background}
    \dd s^2 = -\dd t^2 + a^2(t) \, \dd x^i \dd x_i \, ,
\end{equation}
where $a$ is the scale factor. The background equations of motion are
\begin{equation} \label{eq:Friedmann}
    \ddot{\phi} + 3H\dot{\phi} + V'(\phi) = 0 \, , \qquad
    3H^2 = \frac{1}{2}\dot{\phi}^2 + V(\phi) \, ,
\end{equation}
where $H\equiv \dot{a}/a$ is the Hubble parameter and dot denotes a derivative with respect to the cosmic time $t$. Solutions of these equations are characterized by the slow-roll parameters
\begin{equation} \label{eq:sr_parameters}
    \epsilon_1 \equiv -\frac{1}{H}\frac{\dd \ln H}{\dd t} = \frac{1}{2}\frac{\dot{\phi}^2}{H^2} \, , \qquad
    \epsilon_2 \equiv \frac{1}{H}\frac{\dd \ln \epsilon_1}{\dd t} \, .
\end{equation}
In particular, $\epsilon_1 < 1$ is the sufficient and necessary condition for inflation.

\subsection{Linear perturbations}
\label{sec:linear_perturbations}
Consider next small scalar perturbations around the background,
\begin{align}
    \label{eq:phi_perturbed}
    \phi &= \bar{\phi}(t) + \delta\phi \, , \\
    \label{eq:ds2_perturbed}
    \dd s^2 &= - ( 1 + 2 A ) \dd t^2 + 2 a(t) \partial_i B \dd x^i \dd t + a^2(t) [ ( 1 + 2 \psi ) \delta_{ij} + 2 \partial_i \partial_j E ] \dd x^i \dd x^j \, ,
\end{align}
where $\bar{\phi}$ is the background field, $\delta\phi$ is the field perturbation, $\psi$ is the curvature perturbation, and $A$, $B$, and $E$ are the other metric scalar perturbations, all of which depend both on time and the spatial coordinates. The perturbations are connected by the various components of the perturbed Einstein equation, and to first order, in the spatially flat gauge where $\psi = 0$, they can all be written in terms of $\delta\phi$ alone, following the equation of motion (see, e.g., \cite{Malik:2008im})
\begin{equation} \label{eq:dphi_eom}
    \delta\ddot{\phi}
    + 3H\delta\dot{\phi} + \qty[ -\frac{1}{a^2}\partial^i\partial_i + V''(\bar{\phi}) - \frac{1}{a^3} \frac{\dd}{\dd t}\qty(\frac{a^3}{H}\bar{\phi}^2) ]\delta\phi = 0 \, .
\end{equation}

At this point, I promote the perturbations to a quantum field, described semi-classically as quantum fluctuations over the classical background given by $\bar{\phi}(t)$ and $a(t)$. At linear level, the Fourier modes of $\delta\phi$ are independent harmonic oscillators, with the standard representation\footnote{To be precise, quantization is done in terms of the Sasaki--Mukhanov variable $\nu=a\delta\phi$, whose action reduces to that of Minkowski space quantum field theory in the small-wavelength limit and in conformal time, $\dd \eta = \dd t / a$ \cite{Birrell:1982ix}. Our representation produces the correct canonical commutation relation $\qty[\hat{\nu}(\eta, \vec{x}), \partial_\eta\hat{\nu}(\eta, \vec{y})] = i\delta^{(3)}(\vec{x}-\vec{y})$.}
\begin{gather}
    \label{eq:dphi_operator}
    \delta\hat{\phi}(t, \vec{x})
    = \int \frac{\dd^3 k}{(2\pi)^{3/2}} \hat{\phi}_{\vec{k}}(t) e^{-i\vec{k} \cdot \vec{x}}
    = \int \frac{\dd^3 k}{(2\pi)^{3/2}}
    \qty[\hat{a}_{\vec{k}}\delta\phi_k(t)
    + \hat{a}^\dagger_{-\vec{k}} \delta\phi^*_k(t)] e^{-i\vec{k} \cdot \vec{x}} \, , \\
    \qty[\hat{a}_{\vec{k}}, \hat{a}^\dagger_{\vec{p}}] = \delta^{(3)}(\vec{k} - \vec{p}) \ , \quad \qty[\hat{a}_{\vec{k}}, \hat{a}_{\vec{p}}] = \qty[\hat{a}^\dagger_{\vec{k}}, \hat{a}^\dagger_{\vec{p}}] = 0 \ .
\end{gather}
Here $\hat{a}^{\dagger}_{\vec{k}}$ and $\hat{a}_{\vec{k}}$ are the standard creation and annihilation operators, and all time evolution is captured in the mode functions $\delta\phi_k$. The modes start at sub-Hubble lengths in the Bunch--Davies vacuum state, the state annihilated by all $\hat{a}_{\vec{k}}$ for mode functions obeying \cite{Birrell:1982ix}
\begin{equation} \label{eq:BD_vacuum}
    \delta \phi_k = \frac{1}{\sqrt{2k}a}e^{ik/(aH)} \quad \text{when} \quad k \gg aH \, .
\end{equation}
The expansion of space stretches the modes to super-Hubble scales and amplifies them according to the equation
\begin{equation} \label{eq:dphi_k_eom}
    \delta\ddot{\phi}_k
    + 3H\delta\dot{\phi}_k + \qty[\frac{k^2}{a^2} + V''(\bar{\phi}) - \frac{1}{a^3} \frac{\dd}{\dd t}\qty(\frac{a^3}{H}\dot{\bar{\phi}}^2) ]\delta\phi_k = 0 \, ,
\end{equation}
the Fourier transformation of \eqref{eq:dphi_eom}. The modes get highly excited, quantified by the correlators
\begin{subequations} \label{eq:dphi_k_correlators}
\begin{align}
    \label{eq:dphi_k_correlator}
    \expval{\hat{\phi}^\dagger_{\vec{k}} \hat{\phi}_{\vec{p}}} &= 
    |\delta\phi_k|^2 \delta^{(3)}(\vec{k} - \Vec{p}) \, , \\
    \label{eq:dphi_prime_k_correlator}
    \expval{\dot{\hat{\phi}}^\dagger_{\vec{k}} \dot{\hat{\phi}}_{\vec{p}}} &= 
    |\delta\dot{\phi}_k|^2 \delta^{(3)}(\vec{k} - \Vec{p}) \, , \\
    \label{eq:dphi_dphi_prime_k_correlator}
    \expval{\hat{\phi}^\dagger_{\vec{k}} \dot{\hat{\phi}}_{\vec{p}}} &= 
    \delta\phi_k\delta\dot{\phi}^*_p \delta^{(3)}(\vec{k} - \Vec{p}) \, .
\end{align}
\end{subequations}
Since I treat the perturbations linearly, the corresponding states are Gaussian. At super-Hubble scales, the quantum states are \emph{squeezed} in the $(\delta\phi_k, \delta\phi_k')$ phase space: $\hat{\phi}_\vec{k}$ and $\hat{\phi}'_\vec{k}$ are highly correlated so that one determines the other. The quantum properties of the correlators get suppressed, and the perturbations essentially behave like classical random fields with statistics given by \eqref{eq:dphi_k_correlators}. For more details, see \cite{Grain:2019vnq, Figueroa:2021zah}.

Finally, a gauge transformation gives the (classical) Gaussian comoving curvature perturbation, $\R = \frac{H}{\dot{\phi}}\delta\phi$.

\subsection{Separate universes and stochastic inflation}
\label{sec:separate_universe_picture}
Solving \eqref{eq:dphi_k_eom} yields the perturbations in linear perturbation theory, but as discussed in the introduction, we want access to large, non-linear perturbations. This can be achieved with the separate universe picture \cite{Salopek:1990jq, Wands:2000dp}. Let us, therefore, write the modified Fourier expansions
\begin{subequations}\label{eq:phi_pi_coarse_grained}
\begin{align}
    \label{eq:phi_coarse_grained}
    \phi_R(t,\vec{x}) &\equiv \int \frac{\dd^3 k}{(2\pi)^{3/2}} W(kR) \phi_\vec{k}(t) e^{-i\vec{k} \cdot \vec{x}} \, , \\
    \label{eq:pi_coarse_grained}
    \pi_R(t,\vec{x}) &\equiv \int \frac{\dd^3 k}{(2\pi)^{3/2}} W(kR) \frac{\dot{\phi}_\vec{k}(t)}{H(t)} e^{-i\vec{k} \cdot \vec{x}} \, .
\end{align}
\end{subequations}
These are the coarse-grained field and the field's coarse-grained time derivative, or `momentum'\footnote{This is not the canonical momentum in the sense of classical mechanics, just a convenient piece of notation. Defined this way, $\pi_R$ corresponds to the derivative of the field with respect to the number of e-folds $N$, see below.}.
The window function $W(x)$ goes to $1$ for small $x$ and to $0$ for large $x$. The coarse-grained quantities thus only get contributions from Fourier modes $\phi_\vec{k}$ with wavelengths $\gtrsim R$, the (comoving) coarse-graining length. They represent weighted averages over a patch of size $R$.
I set $R = R_\sigma \equiv 1/(\sigma aH_0)$, where $\sigma \ll 1$ is a constant and $H_0$ is the initial Hubble radius\footnote{$H$ is stochastic, so using it to define the coarse-graining scale would lead to complications. With our convention, used before in, e.g., \cite{Tomberg:2023kli}, the time dependence of $R$ arises only from $a$, whose non-stochasticity I will ensure below. I assume $H$ does not change much during the period of inflation we're interested in, so $R_\sigma$ is always slightly super-Hubble.}. With this choice, $R$ changes in time, always hovering slightly above the Hubble radius. For simplicity, I use a sharp cutoff in Fourier space,
\begin{equation} \label{eq:W_sharp}
    W(x)=
    \begin{cases}
        1, &x < 1 \\
        0, &x \geq 1 \, ,
    \end{cases}
\end{equation}
as is standard in the literature, but the general ideas in this paper also apply to other coarse-graining schemes. I denote the cutoff in $k$-space by $k_R \equiv 1/R$. With the above choice of $R$, this $k_R = k_\sigma \equiv \sigma aH_0$. Later, I will also consider values of $R$ and $k_R$ that differ from $R_\sigma$ and $k_\sigma$.

The time evolution of the coarse-grained quantities has two components.
First, the separate universe picture lets us treat the super-Hubble variables $\phi_R$ and $\pi_R$ like the background quantities of a local FLRW universe, to leading order in the gradient expansion \cite{Salopek:1990jq, Wands:2000dp}. They then follow the Friedmann equations \eqref{eq:Friedmann}. This is encoded in the time dependence of the Fourier modes $\phi_\vec{k}(t)$ (and $H(t)$) in \eqref{eq:phi_pi_coarse_grained}.
Second, as time goes on, the short-wavelength modes initially excluded from $\phi_R$ and $\pi_R$ get stretched to the point where they cross the coarse-graining scale, becoming part of the coarse-grained quantities. This is captured by the time dependence of $R$ in $W(kR)$.
I treat the short-wavelength modes linearly and quantum mechanically, as described in the previous section. As they get integrated into the classical background quantities $\phi_R$ and $\pi_R$, the latter undergo random jumps or `kicks', leading to a stochastic evolution. Instead of a unique time evolution, we get a family of solutions $\phi_R(N)$ corresponding to different realizations of the stochastic noise, from which we can infer the statistical properties of $\phi_R$ such as its probability distribution, mean, and variance.

At super-Hubble scales, the state of the system is uniquely determined by $\phi$, its time derivative, the time $t$, and the scale factor $a$, which I write in terms of the number of e-folds of expansion $N\equiv\ln a$. In principle, all perturbed quantities receive stochastic kicks. It is convenient to switch to $N$ as the time variable, that is, to describe the system at consecutive $N$=constant hypersurfaces, eliminating $N$-perturbations\footnote{This is a gauge choice. Technically, we should now evaluate $\delta\phi$ in the uniform-$N$ gauge instead of the spatially flat gauge used in \eqref{eq:dphi_eom}. However, these two gauges are close to each other at super-Hubble scales, and the spatially flat gauge is easier to work in \cite{Figueroa:2021zah, Pattison:2019hef}.} and making $N$ a convenient classical clock variable. This allows us direct access to $N$ in each stochastic realization, useful to access the curvature perturbation through the $\Delta N$ formalism \cite{Sasaki:1995aw, Sasaki:1998ug, Wands:2000dp, Lyth:2004gb, Fujita:2013cna, Fujita:2014tja, Vennin:2015hra}\footnote{In the process, we lose track of the cosmic time $t$, which in the uniform-$N$ gauge becomes a perturbed, stochastic quantity. This is fine---$t$ is not interesting observationally.}. The $\Delta N$ formalism concerns the boundary conditions of the stochastic evolution. In this paper, we study the bulk stochastic evolution and the $\Delta N$ formalism won't enter explicitly, but it informs our choices.

Combining these considerations, the evolution equations become
\begin{equation} \label{eq:Friedmann_stochastic}
    \phi_R' = \pi_R + \xi_\phi \, ,
    \qquad
    \pi_R' = -\qty(3 - \frac{1}{2}\pi_R^2)\pi_R - \frac{V'(\phi_R)}{H^2} + \xi_\pi \, ,
    \qquad
    H^2 = \frac{V(\phi_R)}{3-\frac{1}{2}\pi_R^2} \, ,
\end{equation}
where prime denotes a derivative with respect to $N$. The quantities $H$ and $N$ here also describe the local background universe, even though I omit the index $R$ for them. The stochastic kicks are described by the $\xi_\phi$ and $\xi_\pi$ terms, which by a direct differentiation of \eqref{eq:phi_pi_coarse_grained} take the form\footnote{I'm slightly abusing the notation here---in \eqref{eq:xi_computed}, $\xi_\phi$ and $\xi_\pi$ are real random variables, while $\hat{\phi}_{\vec{k}}$ are quantum operators in a complex Hilbert space at the verge of exiting the quantum mechanical short-wavelength regime. However, in the squeezed super-Hubble limit, the quantum variables behave in an almost classical way, as explained in the text.}
\begin{subequations}\label{eq:xi_computed}
\begin{align}
    \label{eq:xi_phi_computed}
    \xi_\phi(N) &= \int \frac{d^3 k}{(2\pi)^{3/2}} \frac{\dd W(kR_\sigma)}{\dd N} \hat{\phi}_{\vec{k}}(N) e^{-i\vec{k} \cdot \vec{x}} = \frac{\dd k_\sigma}{\dd N}\int \frac{d^3 k}{(2\pi)^{3/2}} \delta(k - k_\sigma) \hat{\phi}_\vec{k}(N) e^{-i\vec{k} \cdot \vec{x}} \, , \\
    \label{eq:xi_pi_computed}
    \xi_\pi(N) &= \int \frac{d^3 k}{(2\pi)^{3/2}} \frac{\dd W(kR_\sigma)}{\dd N} \hat{\phi}'_{\vec{k}}(N) e^{-i\vec{k} \cdot \vec{x}} = \frac{\dd k_\sigma}{\dd N}\int \frac{d^3 k}{(2\pi)^{3/2}} \delta(k - k_\sigma) \hat{\phi}'_\vec{k}(N) e^{-i\vec{k} \cdot \vec{x}} \, ,
\end{align}
\end{subequations}
where I pulled the Fourier modes at the $R$-boundary from the short-wavelength quantum regime, hence the hats. We can use \eqref{eq:dphi_k_correlators} to compute the two-point functions
\begin{subequations}\label{eq:xi_correlators}
\begin{align}
    \label{eq:correlator_phiphi}
    \expval{\xi_\phi(N)\xi_\phi(N')} &= \frac{1}{6\pi^2}\frac{\dd k_\sigma^3}{\dd N}|\delta\phi_{k_\sigma}(N)|^2\delta(N-N') \, , \\
    \label{eq:correlator_pipi}
    \expval{\xi_\pi(N)\xi_\pi(N')} &= \frac{1}{6\pi^2}\frac{\dd k_\sigma^3}{\dd N}|\delta\phi'_{k_\sigma}(N)|^2\delta(N-N') \, , \\
    \label{eq:correlator_phipi}
    \expval{\xi_\phi(N)\xi_\pi(N')} &= \frac{1}{6\pi^2}\frac{\dd k_\sigma^3}{\dd N}\delta\phi_{k_\sigma}(N)\delta\phi'^*_{k_\sigma}(N)\delta(N-N') \, .
\end{align}
\end{subequations}
In this linear order, we get white Gaussian noise, completely characterized by the correlators \eqref{eq:xi_correlators}. To solve the mode functions, I write \eqref{eq:dphi_k_eom} in terms of $N$:
\begin{equation} \label{eq:dphi_k_eom_N}
    \delta\phi_k'' = -\qty(3 - \frac{1}{2}\pi_R^2)\delta\phi_k' - \qty[\frac{k^2}{a^2H^2} +
    \pi_R^2\qty(3-\frac{1}{2}\pi_R^2) +
    2\pi_R\frac{V'(\phi_R)}{H^2} + \frac{V''(\phi_R)}{H^2}
    ]\delta\phi_k \, ,
\end{equation}
where the classical background equations \eqref{eq:Friedmann} were used to express everything in terms of $\phi_R$, $\pi_R$, and $H$. The latter can further be expressed in terms of $\phi_R$ and $\pi_R$ using the last equation in \eqref{eq:Friedmann_stochastic}.

Note that I used the classical background equations without the noise terms when manipulating \eqref{eq:dphi_k_eom}, and hence, the noises $\xi_\pi$ and $\xi_\phi$ don't appear in \eqref{eq:dphi_k_eom_N}. I will discuss this further in the next section.

In the classical, squeezed limit discussed at the end of Section~\ref{sec:linear_perturbations}, the two noise terms are fully correlated, $\xi_\pi=\xi_\phi \frac{\delta\phi'_{k_\sigma}}{\delta\phi_{k_\sigma}}$ (in this limit, the coefficient of proportionality is real) \cite{Figueroa:2020jkf, Figueroa:2021zah, Tomberg:2022mkt}. To solve equations \eqref{eq:Friedmann_stochastic}, we need to generate only one Gaussian random variable at each time step, determining the stochastic kicks to both $\phi_R$ and $\pi_R$. It is the classicality of the correlators that allows us to treat $\xi_\phi$ and $\xi_\pi$ as consistent noise for the classical field and its momentum.

The equations presented above are well known in the stochastic inflation literature, see, e.g., \cite{Pattison:2019hef, Figueroa:2021zah} for recent work and \cite{Salopek:1990re} for an early similar take. Before moving to novel results, let me emphasize that the local Friedmann equations \eqref{eq:Friedmann_stochastic} are fully non-linear in the field and momentum. They can describe large field perturbations for which linear perturbation theory fails. This is the power of stochastic inflation. 

\subsection{Zoom-in schemes}
\label{sec:alternating_scheme_introduced}
The above derivation of \eqref{eq:Friedmann_stochastic} may seem ad hoc: I simply stated that the evolution splits into two parts, the classical drift and the quantum noise, whose forms were deduced separately and then added together. Equations \eqref{eq:Friedmann_stochastic} are indeed the form utilized in the literature, but their derivation should be made more transparent.

\begin{figure}
    \centering
    \includegraphics{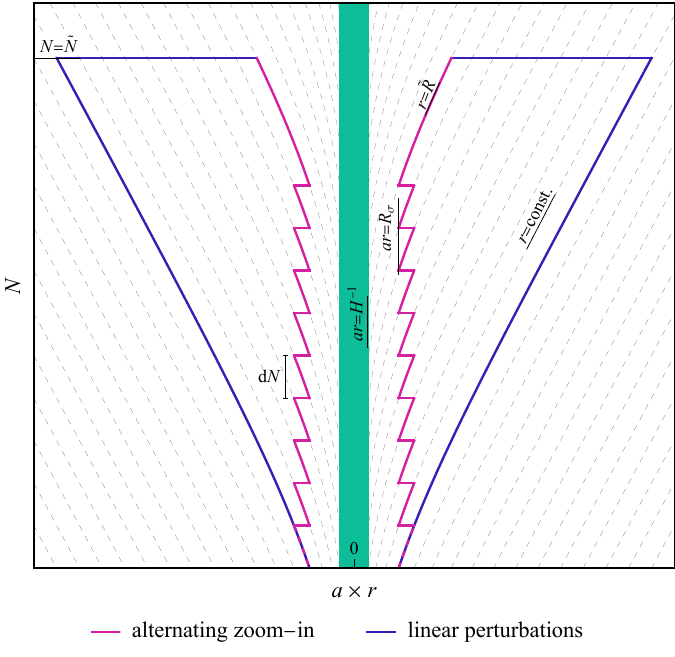}
    \caption{The coarse-graining scale $R$ versus the time $N$ in the alternating zoom-in scheme (magenta) and linear perturbation theory (blue) when computing $\phi_R$ for $R=\tilde{R}$ at time $N=\tilde{N}$. The dashed lines correspond to fixed values of $r$, the comoving radial coordinate in which $R$ is measured; the x-axis depicts the physical distance $ar$. The green region lies inside the Hubble radius.}
    \label{fig:zoom}
\end{figure}

To achieve this, let us turn off the noise terms $\xi_\phi$ and $\xi_\pi$ for a moment, starting from time $N_0$. The variables $\phi_R$ and $\pi_R$ then follow the local FLRW evolution, as given by the gradient expansion. The coarse-graining scale drifts with the Hubble flow, $R = 1/k_R = R_\sigma(N_0) = \text{constant}$, so that no scales cross the barrier and no stochastic noise is induced. Let us continue this time evolution for some finite time $\dd N$. After this time, to compensate, let us bring $R$ back to the current value of $R_\sigma$ in one instantaneous jump. This corresponds to `zooming in' in the comoving coordinates, from the initial $R=R_\sigma(N_0)$ to the subsequent $R=R_\sigma(N_0 + \dd N)$. Based on \eqref{eq:phi_pi_coarse_grained}, this results in macroscopic kicks in $\phi_R$ and $\pi_R$ of magnitudes
\begin{subequations}\label{eq:macro_step}
\begin{align}
    \label{eq:phi_macro_step}
    \dd \phi_R &= \int \frac{\dd^3 k}{(2\pi)^{3/2}} \qty[W(kR_\sigma(N_0 + \dd N)) - W(kR_\sigma(N_0))] \hat{\phi}_\vec{k}(N_0+\dd N) e^{-i\vec{k} \cdot \vec{x}} \, , \\
    \label{eq:pi_macro_step}
    \dd \pi_R &= \int \frac{\dd^3 k}{(2\pi)^{3/2}} \qty[W(kR_\sigma(N_0 + \dd N)) - W(kR_\sigma(N_0))] \hat{\phi}'_\vec{k}(N_0+\dd N) e^{-i\vec{k} \cdot \vec{x}} \, .
\end{align}
\end{subequations}
Changing $R$ happens instantaneously: $N$ does not change during this step. Let us repeat this pattern, alternating periods of classical evolution and instantaneous stochastic kicks. The process is depicted in Figure~\ref{fig:zoom}. I call it the \emph{alternating zoom-in scheme}. In the $\dd N \to 0$ limit, the kicks \eqref{eq:macro_step} obey
\begin{subequations}\label{eq:micro_step}
\begin{align}
    \label{eq:phi_micro_step}
    \frac{\dd \phi_R}{\dd N} &\xrightarrow{\dd N \to 0}  \int \frac{\dd^3 k}{(2\pi)^{3/2}} \frac{\dd W(kR_\sigma(N_0))}{\dd N} \hat{\phi}_\vec{k}(N_0) e^{-i\vec{k} \cdot \vec{x}} = \xi_\phi(N_0) \, , \\
    \label{eq:pi_micro_step}
    \frac{\dd \pi_R}{\dd N} &\xrightarrow{\dd N \to 0}  \int \frac{\dd^3 k}{(2\pi)^{3/2}} \frac{\dd W(kR_\sigma(N_0))}{\dd N} \hat{\phi}'_\vec{k}(N_0) e^{-i\vec{k} \cdot \vec{x}} = \xi_\pi(N_0) \, ,
\end{align}
\end{subequations}
giving a stochastic process that can be described by equations \eqref{eq:Friedmann_stochastic}.

It should now be obvious why the noise is absent from the $\delta \phi_k$ equation \eqref{eq:dphi_k_eom_N}: the short-wavelength modes $\delta \phi_k$ evolve only during the free-flowing step, when time ticks forward and $\phi_R$ and $\pi_R$ behave classically. They are not affected by the zoom-in step.
After the zoom-in, the modes continue to evolve in the new, zoomed-in background.
To make plain the absence of noise in the mode equation, I wrote  \eqref{eq:dphi_k_eom_N} explicitly in terms of $\phi_R$ and $\pi_R$, eliminating, in particular, higher time derivatives. Similarly to the mode equation, the last equation in \eqref{eq:Friedmann_stochastic} (the Friedmann equation) does not include noise terms, since it is simply a constraint between the field evolution and spatial expansion during the classical steps.

The construction of stochastic inflation as alternating periods of classical drift and quantum kicks is not completely novel. In fact, it was alluded to already in the early paper \cite{Salopek:1990re}\footnote{I thank the Referee for pointing this out.}, where the authors considered the stochastic noise as `impulses which continually adjust the initial conditions for the next phase of  drift.' They also considered mode equations that evolve classically in a stochastic background, with classical constraint equations. With the explicit picture of Figure~\ref{fig:zoom}, I aim to sharpen the intuition about this zoom-in process. As we will see next, this picture is useful for comparing different approaches to cosmological perturbation theory and stochastic integration.

If we're interested in perturbations coarse-grained over a certain length scale $R=\tilde{R}$, we may turn off the stochastic process when $R_\sigma$ reaches this value. In other words, we freeze the comoving value of $R$, skipping the zoom-in step in the evolution, so that no new modes exit the coarse-graining scale. The local background then follows the classical background equations \eqref{eq:Friedmann} until a time $N=\tilde{N}$ where the wish to evaluate $\phi_R$ and $\pi_R$ (or, in the $\Delta N$ formalism, until a fixed field value $\phi_R$, where we evaluate the curvature perturbation through $N$). This procedure was used in \cite{Figueroa:2020jkf, Figueroa:2021zah} to obtain the background quantities at the end of inflation, coarse-grained over scales much larger than the value of $R_\sigma$ there. In Figure~\ref{fig:zoom}, this corresponds to the last part of $R$'s evolution.

The zoom-in scheme described here seems natural, but it need not be the only reasonable choice. As a simple but illuminating example, consider a scheme where $\dd N$ is not taken to zero, but instead, we let it stretch to the full duration of inflation. The background evolves classically until $N=\tilde{N}$ at the end of inflation, corresponding to a wavenumber $\tilde{k} = k_\sigma(\tilde{N})$, and all perturbations are obtained in one final, huge stochastic kick, see again Figure~\ref{fig:zoom}. We get
\begin{equation} \label{eq:dphi_linear_stats}
    \dd \phi_R = \int_{k_\text{ini}}^{\tilde{k}} \frac{\dd^3 k}{(2\pi)^{3/2}} \hat{\phi}_\vec{k}(\tilde{N}) e^{-i\vec{k} \cdot \vec{x}} \, , \qquad \expval{\dd \phi_R^2} = \int_{k_\text{ini}}^{\tilde{k}} \frac{k^3}{2\pi^2} |\delta\phi_k(\tilde{N})|^2 \, \dd \ln k \, .
\end{equation}
Now $\dd \phi_R$ follows a Gaussian distribution with the above variance. This is just the standard linear perturbation theory result. (We can further convert $\dd \phi_R$ to the curvature perturbation $\R$ using linear perturbation theory.) This picture illustrates the difference between linear theory and the stochastic equations \eqref{eq:Friedmann_stochastic}: in the latter, we utilize the separate universe approximation more efficiently by keeping $R$ close to the Hubble radius (in the former, it is vastly super-Hubble by $N=\tilde{N}$). This applies the non-linear FLRW evolution to the perturbations, letting us solve them beyond linear order and \eqref{eq:dphi_linear_stats}. Different schemes are different approximations of the system's full evolution, and as such, they may yield different results for the coarse-grained observables. Our task is to find a scheme that is easy enough to implement while yielding accurate enough results.

Even if we maintain the $\dd N \to 0$ limit and let the coarse-graining scale follow $R_\sigma$, stochastic differential equations like \eqref{eq:Friedmann_stochastic} suffer from an ambiguity. As I will discuss in the next section, they have many competing interpretations, varying by the way in which the classical drift and the stochastic kicks are interlaced. Different interpretations lead to different evolutions.
The alternating zoom-in scheme described in this section, on the other hand, is unambiguous, giving a well-defined meaning for \eqref{eq:Friedmann_stochastic}\footnote{That is, assuming it leads to a well-behaved stochastic process in the $\dd N \to 0$ limit. In Section~\ref{sec:markovian}, I show that this is true for Markovian noise. In Section~\ref{sec:non_markovian}, I make the same argument for the general case.}. Next, I will study the relationship between this scheme and the conventional interpretations of \eqref{eq:Friedmann_stochastic} in the limit of Markovian noise.

\section{Markovian noise}
\label{sec:markovian}
In slow-roll inflation, where the slow-roll parameters $\epsilon_1$ and $\epsilon_2$ are small, the stochastic equations \eqref{eq:Friedmann_stochastic} simplify considerably. Classically, the system follows an attractor trajectory in the $(\phi,\pi)$ phase space.
The trajectory can be obtained from \eqref{eq:Friedmann_stochastic} by setting $\pi_R'=\xi_\pi=0$ and taking the $\pi_R^2 \ll 1$ limit, yielding
\begin{equation}\label{eq:sr_attractor}
    \pi_R = -\frac{V'(\phi)}{V(\phi)} \, .
\end{equation}
The stochastic noise terms $\xi_\phi$ and $\xi_\pi$ are also affected by the attractor, and they align to always preserve the relationship \eqref{eq:sr_attractor} \cite{Tomberg:2022mkt}\footnote{The same happens in other attractor setups as well, such as constant-roll inflation \cite{Tomberg:2023kli}.}. It is then enough to study the equation for the remaining field $\phi_R$. To obtain the $\xi_\phi$ noise, we solve the perturbation equation \eqref{eq:dphi_k_eom_N} in slow roll for the standard result
\begin{equation}\label{eq:sr_mode_functions}
    |\delta\phi_k| = \frac{H}{\sqrt{2}k^{3/2}}  \quad \implies \quad \expval{\xi_\phi(N)\xi_\phi(N')} = \frac{H^2}{4\pi^2} \delta(N - N') \, ,
\end{equation}
where the Hubble parameter depends only on the field value through $3H^2 = V(\phi_R)$.
The stochastic equation is then
\begin{equation}\label{eq:sr_phi_stochastic}
    \phi'_R = \underbrace{-\frac{V'(\phi_R)}{V(\phi_R)}}_{\equiv\mu(\phi_R)} + \underbrace{\frac{H(\phi_R)}{2\pi}}_{\equiv\sigma(\phi_R)}\xi(N) \, , \qquad \expval{\xi(N)\xi(N')} = \delta(N-N') \, .
\end{equation}
This is the most common formulation of stochastic inflation in the literature, see, e.g., \cite{Starobinsky:1986fx, Vennin:2015hra}. The system is \emph{Markovian}: the noise only depends on the system's current state \cite{Lawler:2014}. For future convenience, I introduced a rescaled noise variable $\xi(N)$ with a simple normalization and the notations $\mu$ and $\sigma$ for the drift and diffusion strength.

\subsection{\Ito\ versus Stratonovich}
\label{sec:ito_vs_stratonovich}
How is \eqref{eq:sr_phi_stochastic} solved? Let us write down an algorithm to do this time step by time step, with a step length $\dd N$, which we will take to zero in the end. To simplify the notation, I use the functions $\mu$ and $\sigma$ introduced in \eqref{eq:sr_phi_stochastic} and drop the index $R$ from $\phi$. Using the Euler method \cite{Kloeden:1992},
\begin{equation} \label{eq:phi_Ito_step}
    \text{(\Ito)} \qquad \phi(\Nplus) = \phi(N) + \mu[\phi(N)]\dd N + \sigma[\phi(N)]\sqrt{\dd N} \, \xi_N \, , \quad
    \Nplus \equiv N + \dd N \, .
\end{equation}
The discrete noise $\xi_N$ here is normalized to unit variance:
\begin{equation} \label{eq:xi_discrete}
    \expval{\xi_N\xi_{N'}} = \delta_{NN'} \, .
\end{equation}
Equation \eqref{eq:phi_Ito_step} is the \emph{\Ito\ approach} to solving SDEs \cite{Ito:1944}: evaluate the quantities on the r.h.s. of the equation at the beginning of the time step, and use them to obtain $\phi$ at the end of the step.

For ordinary differential equations, Euler's method is not the best choice due to its slow convergence. Instead, we might want to use something like the following midpoint method,
\begin{equation} \label{eq:phi_Stratonovich_step}
\begin{aligned}
    \text{(Stratonovich)} \qquad \phi(\Nplus) = \phi(N)
    &+
    \frac{1}{2}\qty{\mu[\phi(N)]+\mu[\phi(\Nplus)]}\dd N \\
    &+
    \frac{1}{2}\qty{\sigma[\phi(N)]+\sigma[\phi(\Nplus)]}\sqrt{\dd N} \, \xi_N \, . \phantom{\quad\quad\quad}
\end{aligned}
\end{equation}
In the context of SDEs, this is the \emph{Stratonovich approach} \cite{Stratonovich:1966}. Note that \eqref{eq:phi_Stratonovich_step} is an implicit equation for $\phi(\Nplus)$: it is featured on both sides.

\begin{figure}
    \centering
    \includegraphics{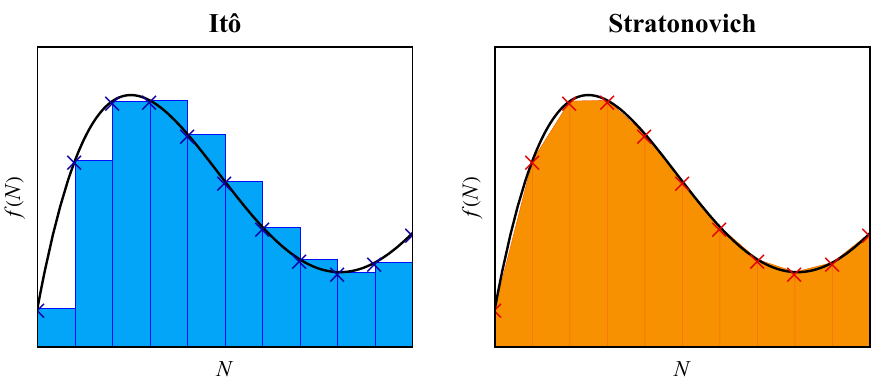}
    \caption{Integration in the \Ito\ versus Stratonovich schemes.}
    \label{fig:schema_integral_bin}
\end{figure}

For ordinary differential equations, both approaches lead to the same final solution for $\phi(N)$. The approaches are related to different methods of numerical integration.
Solving a first-order differential equation of the form $\phi' = f(\phi, N)$ is equivalent to finding $\phi(N)$ that is the $N$-integral of $f(\phi(N),N)$. When computing the integral in the \Ito\ approach, $f$ is approximated in each $N$-bin by its value at the bin's starting point. In the Stratonovich approach, the integral is instead evaluated using the trapezoidal rule. See Figure~\ref{fig:schema_integral_bin} for a comparison between the methods. For non-stochastic variables, both methods converge to the same result when $\dd N \to 0$. However, this is no longer true for stochastic variables.

The problem in the stochastic case arises from the noise term in \eqref{eq:sr_phi_stochastic}, scaling as $\sqrt{\dd N}$. As $\dd N \to 0$, this term naively seems to dominate over the drift, which is proportional to $\dd N$. In truth, we need to keep track of both contributions. Roughly speaking, the deterministic $\mathcal{O}(\dd N)$ terms affect the mean of the $\phi$ distribution, while the stochastic $\mathcal{O}(\sqrt{\dd N})$ terms affect the distribution's width. However, sometimes these terms mix.

Let us solve for $\phi(\Nplus)$ for the Stratonovich step \eqref{eq:phi_Stratonovich_step} keeping only terms up to $\mathcal{O}(\dd N)$. For this, I Taylor expand $\mu(\phi(\Nplus)) = \mu(\phi(N)) + \mu'(\phi(N))\Delta \phi + \dots$ and $\sigma(\phi(\Nplus)) = \sigma(\phi(N)) + \sigma'(\phi(N))\Delta \phi + \dots$. Here $\Delta\phi = \phi(\Nplus) - \phi(N)$ has terms of order $\sqrt{\dd N}$ and higher. Since the $\mu$ term in \eqref{eq:phi_Stratonovich_step} already has an explicit $\dd N$ factor, only the leading term contributes in the $\mu$ expansion. In the $\sigma$ expansion, on the other hand, we need to keep contributions up to $\mathcal{O}(\sqrt{\dd N})$, since these can combine with the explicit $\sqrt{\dd N}$ factor into a term of order $\dd N$. Treating \eqref{eq:phi_Stratonovich_step} iteratively, the only $\sqrt{\dd N}$ contribution to $\Delta \phi$ is $\sigma(\phi(N))\sqrt{\dd N}\xi_N$.
The corresponding term ends up with a factor of $\xi_N^2$, which we can replace by its expectation value of $1$ (for details, see Appendix~\ref{sec:stochastic_formulas}).
We thus get, up to $\mathcal{O}(\dd N)$, \cite{Stratonovich:1966}
\begin{equation} \label{eq:phi_Stratonovich_itoed_step}
\begin{aligned}
    \text{(Stratonovich A)} \qquad \phi(\Nplus) = \phi(N)
    &+
    \qty( \mu[\phi(N)] + \frac{1}{2}\sigma[\phi(N)]\sigma'[\phi(N)]) \dd N \\
    &+ \sigma[\phi(N)] \sqrt{\dd N} \, \xi_N \, .
\end{aligned}
\end{equation}
This equation is of the \Ito\ form \eqref{eq:phi_Ito_step}, but with the added drift $\frac{\sigma}{2}\frac{\partial\sigma}{\partial\phi}$. The additional term is non-zero for \emph{multiplicative noise}, that is, if the $\sigma$ factor multiplying $\xi$ in \eqref{eq:sr_phi_stochastic} depends on $\phi$. Clearly, this changes the solution.

For future reference, I mention here another implementation of the Stratonovich approach, the Euler--Heun method \cite{Foster:2020, Kloeden:1992}. In it, we first obtain an approximative $\phi(\Nplus)$ using the \Ito\ step \eqref{eq:phi_Ito_step}, and then we substitute this into the r.h.s. of \eqref{eq:phi_Stratonovich_step}:
\begin{equation} \label{eq:euler_heun}
    \text{(Stratonovich B)} \qquad \phi(\Nplus) = \qty[\text{r.h.s. of \eqref{eq:phi_Stratonovich_step} with $\phi(\Nplus)$ from \eqref{eq:phi_Ito_step}}] \, .
\end{equation}
Expanding in powers of $\dd N$ produces \eqref{eq:phi_Stratonovich_itoed_step}, proving \eqref{eq:euler_heun} describes the same stochastic process, but \eqref{eq:euler_heun} is closer in spirit to the original formula \eqref{eq:phi_Stratonovich_step} and contains an explicit product of two $\xi$ noise terms.

The Stratonovich approach can also be interpreted as evaluating the functions $\mu$ and $\sigma$ at the midpoint between $N$ and $\Nplus$ when calculating the $\phi$ step. Such an approach may seem nonsensical since we don't have access to $\phi$ at this midpoint in our stepwise process, but a $\dd N$ expansion of such a scheme again correctly produces \eqref{eq:phi_Stratonovich_itoed_step}.

We have seen that the same continuum equation \eqref{eq:sr_mode_functions} corresponds to two different stochastic processes, depending on its stepwise interpretation. The stochastic equation alone is not a complete description of the system: one also has to provide the interpretation. In fact, there's a continuous infinity of different interpretations, obtained by changing the weights of the coefficients in \eqref{eq:phi_Stratonovich_step} \cite{Mannella:2012}, and one may be able to cook up others. Out of these, the \Ito\ approach is uniquely simple: it gives the step in $\phi$ explicitly, and its numerical implementation is straightforward. However, as I explain in Appendix~\ref{sec:stochastic_formulas}, the \Ito\ approach violates the chain rule: for a smooth function $f$, the time derivative $\frac{\dd}{\dd N} f(\phi)$ is not given by $f'(\phi)\phi'(N)$ but receives an extra contribution, again due to two $\mathcal{O}(\sqrt{\dd N})$ terms combining into a $\mathcal{O}(\dd N)$ term. The Stratonovich approach, on the other hand, turns out to respect the chain rule. Sometimes, the Stratonovich approach is favored in physics, since it mixes the different time steps in its noise term, resembling the limit of a more realistic process with colored noise. However, such arguments are heuristic and depend on the way in which the limit is taken \cite{Kampen:1981}.
It is also noteworthy that the \Ito\ and Stratonovich approaches only differ by a drift term, so a process defined in one can always be converted to another by adding this term, without changing the statistics of $\phi$.

In any case, we don't need to resort to heuristics: we have a full description of the stochastic process, given by our zoom-in scheme.

\subsection{Comparison to the alternating scheme}
\label{sec:alternating_scheme_ito_v_strato}
For our alternating zoom-in scheme, one time step reads:
\begin{equation} \label{eq:phi_alternating_step}
\begin{aligned}
    \text{(Alternating)} \qquad \tilde{\phi}(\Nplus) &= \phi(N) + \mu[\phi(N)]\dd N \, , \\
    \phi(\Nplus) &=  \tilde{\phi}(\Nplus) + \sigma[\tilde{\phi}(\Nplus)]\sqrt{\dd N} \, \xi_N \, . \phantom{\qquad\text{(Alternating)}}
\end{aligned}
\end{equation}
In other words, we first evolve the field with the drift only to the value $\tilde{\phi}(\Nplus)$, and then add the stochastic kick using $\tilde{\phi}(\Nplus)$ as the instantaneous field value. Since $\tilde{\phi}(\Nplus)$ doesn't contain $\mathcal{O}(\sqrt{\dd N})$ terms, expanding up to $\mathcal{O}(\dd N)$ gives simply
\begin{equation} \label{eq:alternating_itoed_step}
    \qquad \phi(\Nplus) = \phi(N) + \mu[\phi(N)]\dd N + \sigma[\phi(N)]\sqrt{\dd N} \, \xi_N \, .
\end{equation}
In other words, the alternating scheme matches the \Ito\ approach. 
This is one of the novel results of this paper: the \Ito\ approach matches a simple microphysical description of stochastic inflation, while the Stratonovich approach does not. 

Physically, the strength of the stochastic kick is determined by the perturbations exiting the coarse-graining scale.
In the \Ito\ and alternating schemes, the kick strength---and thus the perturbations---only depend on the pre-kick field value $\phi(N)$, that is, on evolution up until the kick.
In the Stratonovich scheme, the kick strength depends also on the post-kick field value $\phi(\Nplus)$, that is, the kicked field seems to affect the prior evolution of the perturbations in an acausal manner.

\subsection{How big is the difference?}
\label{sec:how_big_difference}
As we saw above, the difference between the \Ito\ and Stratonovich approaches is an extra drift term of the form $\frac{\sigma}{2}\frac{\partial\sigma}{\partial\phi}$. This should be compared to the original drift $\mu$. For transparency, I restore the units of the reduced Planck mass $\MPl$ for this section and the next, so that the $\phi$ dependence of $H$ is given by $3\MPl^2 H^2 = V(\phi)$. We then have
\begin{align}
    \label{eq:mu_planck}
    \mu(\phi) &= -\frac{V'(\phi)}{3H^2(\phi)} = -\frac{V'(\phi)\MPl^2}{V(\phi)} \, , \\
    \label{eq:extra_drift_planck}
    \frac{\sigma(\phi)}{2}\frac{\partial \sigma(\phi)}{\partial\phi}
    &= \frac{1}{4}\frac{\partial \sigma^2(\phi)}{\partial\phi}
    = \frac{1}{16\pi^2}\frac{\partial H^2(\phi)}{\partial\phi}
    = \frac{V'(\phi)}{48\pi^2\MPl^2} \, .
\end{align}
The extra drift always points in the opposite direction from $\mu$: it pushes the field distribution towards larger potential values, where the diffusion coefficient $\sigma$ is larger. For the extra term to be negligible, we need
\begin{equation} \label{eq:mu_comparison}
    \left|\frac{\sigma}{2}\frac{\partial \sigma}{\partial \phi}\right| \ll |\mu| \quad \iff \quad H \ll 4\pi \MPl \quad \iff \quad V \ll 48\pi^2\MPl^4 \, .
\end{equation}
The difference is suppressed by $H/\MPl$. This is not a new result: it was already noted in \cite{Salopek:1990re}, and later in \cite{Pinol:2018euk}.
Observable inflation takes place at sub-Planckian energies \cite{Planck:2018jri}, where the \Ito\ and Stratonovich approaches agree with each other. Depending on the model, the difference may be important for eternal inflation \cite{Goncharov:1987ir, Guth:2000ka, Linde:2015edk}, where the field may probe Planckian energy densities \cite{Linde:1986fc}, although the semi-classical formalism arguably breaks down in this regime as quantum gravity becomes important.
The difference between the \Ito\ and Stratonovich approaches thus goes beyond the applicability of the leading-order stochastic slow-roll equation \eqref{eq:sr_phi_stochastic}, as noted before in \cite{Winitzki:1995pg, Vennin:2015hra}. 
Since the limit \eqref{eq:mu_comparison} was derived from \eqref{eq:sr_phi_stochastic}, it is possible (though maybe not likely) that deviations from slow roll, such as ultra slow roll \cite{Dimopoulos:2017ged}, may yield observable differences in different stochastic schemes.

\subsection{Numerical example}
To demonstrate the concepts of the previous sections, let us study numerically a simple model with
\begin{equation} \label{eq:numerical_model}
    V(\phi) = \frac{1}{2}m^2\phi^2 \quad \implies \quad \mu(\phi) = -\frac{2\MPl^2}{\phi} \, , \quad \sigma(\phi) = \frac{m\phi}{2\sqrt{6}\pi\MPl} \, .
\end{equation}
Using the parameter values $m=0.1\MPl$, $0.5\MPl$, and $1\MPl$, I evolved $\phi$ for 100 e-folds, starting from $\phi=50\MPl$ (so that $\frac{H}{4\pi\MPl}=0.16$, $0.81$, and $1.62$ initially in the three cases), with four numerical schemes: the \Ito\ approach \eqref{eq:phi_Ito_step}, the alternating zoom-in scheme \eqref{eq:phi_alternating_step}, the Stratonovich approach transformed into the \Ito-like form (Stratonovich A) \eqref{eq:phi_Stratonovich_itoed_step}, and the Stratonovich approach with the Euler--Heun method (Stratonovich B) \eqref{eq:euler_heun}. There is an absorbing boundary at $\phi=\sqrt{2}\MPl$, where the first slow-roll parameter crosses one and inflation ends. The super-Planckian parameters are chosen to produce noticeable differences between the \Ito\ and Stratonovich approaches, as discussed in Section~\ref{sec:how_big_difference}. I ran $10^7$ realizations of the stochastic process for each case and collected the statistics.

\begin{figure}[p]
    \centering
    \includegraphics{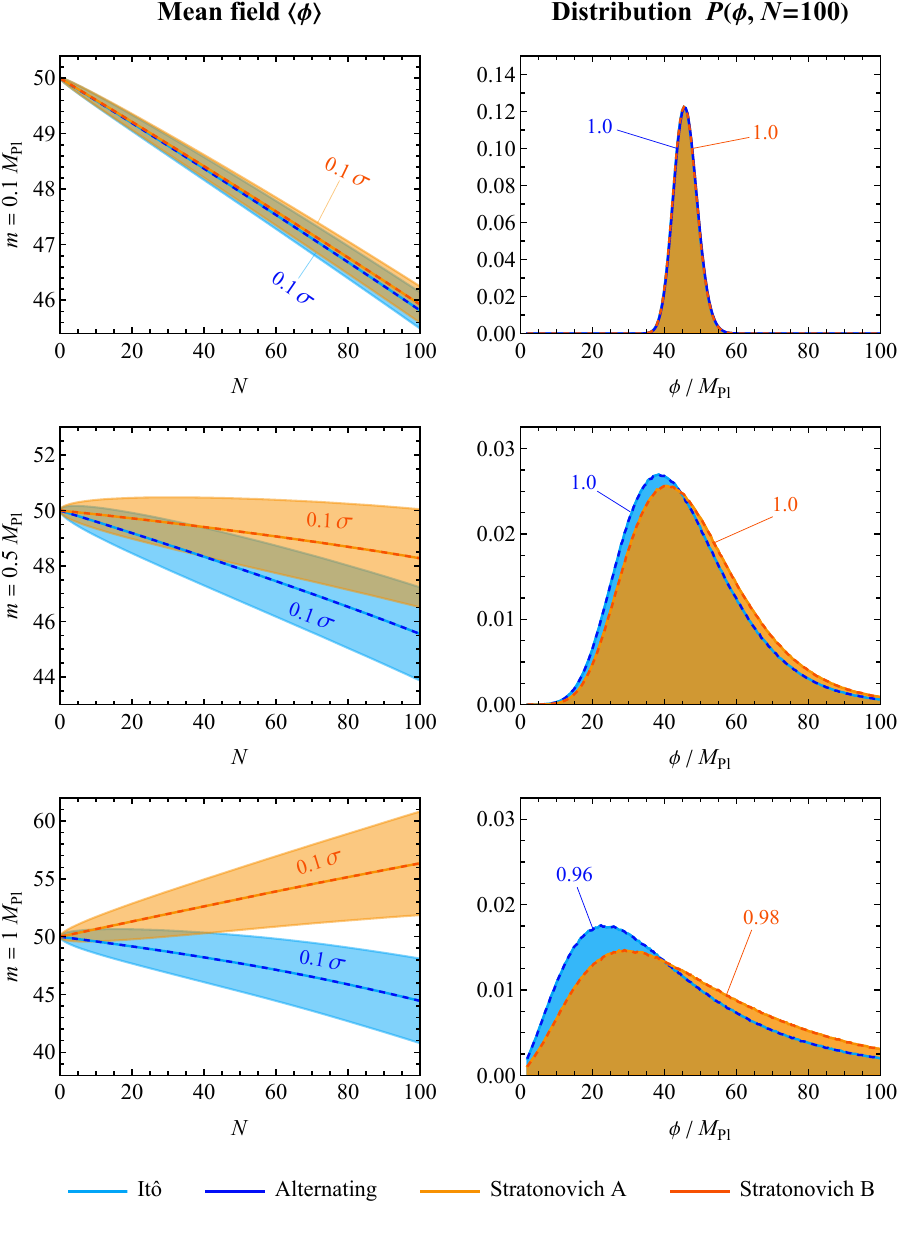}
    \caption{Evolution of the probability distribution in the $\phi^2$ model with different values of $m$, in the different stochastic schemes. \emph{Left:} The mean field value in terms of e-folds. The colored region covers $\pm 0.1$ standard deviations around the mean. \emph{Right:} The field's probability distribution at $N=100$. The numbers indicate the fraction with $\phi > \sqrt{2}\MPl$, that is, the fraction not absorbed into the boundary.}
    \label{fig:phi2_comparison}
\end{figure}

The results are depicted in Figure~\ref{fig:phi2_comparison}. The figure shows the time evolutions of the mean field value and its variance and the full $\phi$ distribution at the final time. In all cases, the field starts with the fixed initial value with no variance, but as time goes on, the variance grows and the field distribution spreads. For large $m$ and thus $\sigma$, the field spreads fast to large values, but an increasing portion also gets absorbed to the boundary.

As expected based on the above discussion, the \Ito\ and alternating schemes agree with each other to an excellent degree, and similarly, the two Stratonovich schemes agree with each other. The Stratonovich and \Ito\ distributions differ more and more for increasing $m$. When $m$ (and thus the typical energy density) is sub-Planckian, the difference is negligible, and the classical drift $\mu$ pulls the mean field value towards zero. When $m$ grows, the extra drift of the Stratonovich approach starts to oppose $\mu$ and overcomes it around the Planck scale. In this limit, the mean field value actually grows in the Stratonovich approach. Interestingly, the peak of the probability distribution still decreases in time---the growth of the mean is related to the ever-widening tail. Similar results were earlier obtained in \cite{Salopek:1990re}, see their Figure~4.

\subsection{Validity of the Markovian approximation}
\label{sec:validity_of_markovian}
Above, I argued the Markovian description of the stochastic process---that is, noise that only depends on the current state of the system---applies in slow roll and possibly in other attractor regimes. For the noise to behave this predictably, the background evolution must be considerably regular, even in the presence of stochasticity. Attractor behavior helps with this, confining the stochastic motion to some degree \cite{Tomberg:2022mkt}, but it alone may not guarantee Markovianity. Let us examine these issues in more detail in slow roll.

The slow-roll form of the noise, \eqref{eq:sr_mode_functions}, depends on the current value of the Hubble parameter $H$.
In fact, the form \eqref{eq:sr_mode_functions} was computed assuming $H$ is approximately constant through the Fourier mode's evolution from the sub-Hubble vacuum to the super-Hubble coarse-graining scale.
This corresponds to the zeroth order of the slow-roll approximation. To account for a time-dependent $H$, it is better to work in terms of the comoving curvature perturbation $\R = \frac{\delta\phi}{\phi'} = \frac{\delta\phi}{\sqrt{2\epsilon_1}}$. It can be shown that, to leading order in slow roll, $\R_k$ freezes to a constant value $H/(2\sqrt{\epsilon_1 k^3})$ at super-Hubble scales, where $H$ and $\epsilon_1$ are evaluated at the mode's Hubble exit (see, e.g., \cite{Lyth:2009zz}). Note the discrepancy between the time when the functions are evaluated (Hubble exit, $aH=k$) and the time they give the stochastic kick (coarse graining, $\sigma aH=k$). This discrepancy is necessary: we want the coarse-graining scale to be sufficiently super-Hubble so that the gradient expansion applies and the modes have assumed their super-Hubble values and are squeezed enough for the correlators to become classical. Reversing the transformation between $\R$ and $\delta\phi$, we get the corrected coarse-graining mode functions
\begin{equation} \label{eq:corrected_sr_mode_functions}
    |\delta\phi_k(N_\text{coarse})| = \frac{H(N_\text{coarse})}{\sqrt{2}k^{3/2}} \times
    \sqrt{\frac{\epsilon_1(N_\text{coarse})}{\epsilon_1(N_\text{H-exit})}} \frac{H(N_\text{H-exit})}{H(N_\text{coarse})} \, .
\end{equation}
From the definitions of the slow-roll parameters, $H(N+\Delta N)/H(N) \approx 1 - \epsilon_1(N)\Delta N$ and $\epsilon_1(N+\Delta N)/\epsilon_1(N) \approx 1 + \epsilon_2(N)\Delta N$. For a typical value of $\sigma \sim 0.01$ \cite{Figueroa:2021zah}, we have $\Delta N \approx \ln \sigma \approx 5$ between the Hubble exit and coarse graining, so, in the slow-roll limit of $\epsilon \ll 1$, $\epsilon_2 \ll 1$, the difference between \eqref{eq:corrected_sr_mode_functions} and \eqref{eq:sr_mode_functions} is small. For slow-roll parameters closer to $1$, it may become significant.
Higher slow-roll orders bring more corrections.

Result \eqref{eq:corrected_sr_mode_functions} is no longer Markovian: it depends on the evolution of the field prior to $N_\text{coarse}$. It also depends on the coarse-graining parameter $\sigma$, which was absent from the leading result \eqref{eq:sr_mode_functions}.
Moreover, \eqref{eq:corrected_sr_mode_functions} becomes unreliable in a stochastic background, since it was derived in the classical background, and the stochastic time-evolution deviates from the classical slow-roll behavior. (Due to the attractor nature of slow roll, the stochastic motion still happens along the classical trajectory in the $(\phi,\pi)$ phase space, as discussed above and in \cite{Tomberg:2022mkt}, but the time dependence of $\phi$, and thus of $H$, may differ from the classical one.) The simple result \eqref{eq:sr_mode_functions} is likely the best Markovian approximation for the noise in the regime of small to mid-sized perturbations, even though it suffers from the ambiguity related to the difference between the Hubble and coarse-graining scales.
This ambiguity affects the comparison between the \Ito\ and Stratonovich approaches since their difference is related to the noise's $\phi$ dependence.
If the stochastic noise is extremely strong (either for the model in general, or for a rare stochastic realization), corresponsing to large perturbations, $H$ may change so quickly during a single mode's evolution that the Markovian approximation breaks down completely. 
 
Besides slow roll, an interesting example is constant-roll inflation near a local maximum (or minimum) of the potential \cite{Tomberg:2023kli}. The second slow-roll parameter $\epsilon_2$ is a constant there; it may be large, but even then, constant roll is an attractor, so the field is confined to its classical trajectory. Furthermore, the energy scale is bound by the extremum of the potential, making $H$ approximately a constant. After constant roll has continued for a while, a result similar to \eqref{eq:sr_mode_functions} applies to the mode functions.

In fact, constant-roll hilltop inflation is arguably the only limit in canonical single-field inflation where the Markovian approximation is exact. For such exactness, the mode equation \eqref{eq:dphi_k_eom_N} must have no dependence on $\phi_R$ or $\pi_R$, since a dependence would inevitably make the final $\delta\phi_k$ depend on the prior stochastic evolution. In particular, this requirement applies to $H$, which must then be close to a constant, implying $\pi_R = \sqrt{-2\partial_N \ln H} \approx 0$. Only the $V''(\phi_R)/H^2$ term remains in \eqref{eq:dphi_k_eom_N}, so we finally need $V''(\phi_R)=\text{constant}$. This results in constant-roll inflation close to a potential minimum or maximum, where the constant $V''(\phi_R)$ sets the constant value of $\epsilon_2$, see \cite{Karam:2022nym, Tomberg:2023kli}. All modes then evolve identically (after constant roll starts), and the stochastic noise coefficient $\sigma$ retains no dependence on $\phi_R$. The \Ito\ and Stratonovich approaches are thus equivalent.
Even then, strong enough stochastic kicks will drive the field away from the hilltop, where the Markovian approximation needs to be re-examined.

Beyond inflationary attractors, e.g., during ultra slow roll \cite{Pattison:2021oen}, the Markovian approximation becomes even worse, and, for accurate results, one must solve the full equations \eqref{eq:Friedmann_stochastic} instead. I will discuss this in the next section. In such scenarios, $\epsilon_2$ is typically strongly negative or changes quickly in time.

\section{Beyond Markovian noise}
\label{sec:non_markovian}
The full set of stochastic equations \eqref{eq:Friedmann_stochastic}, \eqref{eq:xi_correlators}, \eqref{eq:dphi_k_eom_N} can be solved numerically---in the context of primordial black holes, such a program was carried out in \cite{Figueroa:2020jkf, Figueroa:2021zah}. During the time evolution, we need to keep track of $\phi_R$, $\pi_R$, and a number of Fourier modes $\delta\phi_k$, whose evolution we need to follow from sub-Hubble (when they start to deviate from the Bunch--Davies vacuum \eqref{eq:BD_vacuum}) up to the coarse-graining scale. These are all coupled: the $\delta\phi_k$ with $k=k_\sigma$\footnote{In a numerical setup with a discrete number of Fourier modes, this is the mode closest to $k_\sigma$, or an interpolation between modes.} affects the stochastic kicks to $\phi_R$ and $\pi_R$, and $\phi_R$ and $\pi_R$ backreact on  $\delta\phi_k$, affecting the evolution of all modes. Beyond simple cases such as slow roll, the backreaction may be important in capturing all non-Gaussianity in a general case, as argued in \cite{Cruces:2018cvq, Cruces:2021iwq, Cruces:2022imf}. It renders the evolution non-Markovian.

The alternating zoom-in scheme of Section~\ref{sec:alternating_scheme_introduced} can be applied as-is also in the full non-Markovian case. In a numerical setup, we can improve the computation of the classical evolution step of $\phi_R$, $\pi_R$, and $\delta\phi_k$ by choosing, e.g., a high-order Runge--Kutta method, as done in \cite{Figueroa:2020jkf, Figueroa:2021zah}, instead of the Euler method of \eqref{eq:phi_Ito_step}. This choice is independent of the stochastic noise, which is applied between such steps\footnote{This excludes numerical methods that mix different time steps, such as leapfrog integration (see, e.g., \cite{Figueroa:2020rrl}).}. On the other hand, the concept of the \Ito\ and Stratonovich approaches is more involved in the non-Markovian case, as we will see next.

\subsection{\Ito\ versus Stratonovich in the non-Markovian case}
The \Ito\ and Stratonovich approaches are defined for Markovian SDEs of the form \eqref{eq:sr_phi_stochastic}, where the noise depended on the stochastic variable $\phi_R$\footnote{The generalization to a two-dimensional Markovian case where $\pi_R$ is independent of $\phi_R$ and the noise depends on both is simple, although such cases don't tend to crop up in the literature on stochastic inflation.}. In the non-Markovian case, the noise instead depends on $\delta\phi_k$, and the dependence on the stochastic variables $\phi_R$ and $\pi_R$ is indirect, going through the backreaction channel.

To apply the \Ito\ and Stratonovich approaches in the non-Markovian case, I promote the modes $\delta\phi_k$ to equal status with $\phi_R$ and $\pi_R$.
To make the $\delta\phi_k$ equation first order, I also introduce the auxiliary `momentum' variable $\delta\pi_k = \delta\phi_k'$. The full system can be written as
\begin{equation} \label{eq:general_markovian_form}
    \Phi_i' = \mu_i(\Phi_j,N) + \sigma_i^{\alpha}(\Phi_j,N)\xi_\alpha(N) \, , \qquad
    \expval{\xi_\alpha(N)\xi_\beta(N')} = \delta_{\alpha\beta}\delta(N-N') \, ,
\end{equation}
where I gathered all the variables into one vector with components $\Phi_i$\footnote{The notation $\Phi_j$ in the function arguments refers to the collection of all components of the vector $\Phi$.}, whose evolution is governed by the drift vector $\mu_i$ and the diffusion matrix $\sigma_i^\alpha$. From \eqref{eq:Friedmann_stochastic}, \eqref{eq:dphi_k_eom_N}, the vectors have the forms
\begin{equation} \label{eq:general_markovian_form_unpacked}
    \Phi_i =
        \begin{bmatrix}
            \phi_R \\
            \pi_R \\
            \delta\phi_k \\
            \delta\pi_k
        \end{bmatrix} \, , \quad
    \mu_i =
        \begin{bmatrix}
            \pi_R \\
            \mu_{\pi_R}(\phi_R,\pi_R) \\
            \delta\pi_k \\
            \mu_{\delta\pi_k}(\delta\phi_k, \delta\pi_k, \phi_R, \pi_R, N)
        \end{bmatrix} \, , \quad
    \sigma^\alpha_i = 
        \begin{bmatrix}
            \sigma^\alpha_{\delta\phi_R}(\delta\phi_k,\delta\pi_k,N) \\
            \sigma^\alpha_{\delta\pi_R}(\delta\phi_k,\delta\pi_k,N) \\
            0 \\
            0
        \end{bmatrix} \, ,
\end{equation}
where I have indicated the variable dependence of the component functions. The mode functions $\delta\phi_k$, $\delta\pi_k$ represent the set of all the relevant $k$ modes.

The arguments below don't depend on the exact forms of $\mu_i$ and $\sigma_i^\alpha$ beyond the general shapes of \eqref{eq:general_markovian_form_unpacked}, but for completeness, equations \eqref{eq:Friedmann_stochastic}, \eqref{eq:dphi_k_eom_N} give
\begin{equation} \label{eq:unpacked_drifts}
\begin{aligned}
    \mu_{\pi_R} &= -\qty(3-\frac{1}{2}\pi_R^2)\qty(\pi_R + \frac{V'(\phi_R)}{V(\phi_R)}) \, , \\
    \mu_{\delta\pi_k} &= -\qty(3 - \frac{1}{2}\pi_R^2)\qty(\delta\pi_k + \pi_R^2\delta\phi_k + \frac{1}{V(\phi_R)}\qty[\frac{k^2}{a(N)^2} + 2\pi_R V'(\phi_R) + V''(\phi_R)]\delta\phi_k)
\end{aligned}
\end{equation}
for the drift. When it comes to the diffusion strength, in \eqref{eq:general_markovian_form}, I allowed the inclusion of many independent noises $\xi_\alpha$ for generality; in the squeezed limit discussed in Section~\ref{sec:stochastic}, there is only one $\xi(N)$, with
\begin{equation} \label{eq:unpacked_squeezed_sigmas}
\begin{aligned}
    \sigma_{\delta\phi_R} = \sqrt{\frac{k_\sigma^3}{2\pi^2}}|\delta\phi_{k_\sigma}(N)| \, , \qquad 
    \sigma_{\delta\pi_R} =\frac{\delta\pi_{k_\sigma}}{\delta\phi_{k_\sigma}}\sigma_{\delta\phi_R} \, .
\end{aligned}
\end{equation}
These depend on $N$ through $k_\sigma$, which selects the coarse-grained mode for each moment in time. Note that nearby modes are close to each other since their initial conditions \eqref{eq:BD_vacuum} are smooth in $k$ and they share the same background evolution. In other words, $\delta\phi_k(N)$ and $\delta\pi_k(N)$ are continuous functions of $k$.

Going to discrete time steps, equation \eqref{eq:general_markovian_form} can now be interpreted through the different approaches. Analogously to \eqref{eq:phi_Ito_step}, the \Ito\ approaches reads
\begin{equation} \label{eq:general_ito_step}
\begin{aligned}
    \text{(\Ito)} \qquad &\Phi_i(N_+) = \Phi_i(N) + \mu_i[\Phi_j(N),N]\dd N + \sigma_i^{\alpha}[\Phi_j(N),N]\sqrt{\dd N}\xi_{\alpha,N} \, , \phantom{\text{(\Ito)} \qquad} \\
    & \Nplus \equiv N + \dd N \, , \quad
    \expval{\xi_{\alpha,N} \, \xi_{\beta,N'}} = \delta_{\alpha\beta}\delta_{NN'} \, ,
\end{aligned}
\end{equation}
with a straightforward interpretation: we evaluate the functions $\Phi_i$ at $N$ and use the results to update $\Phi_i$ to $\Nplus$.\footnote{For discrete time steps, it is enough to also consider a discrete set of modes $k$, exiting at the discrete times $N$.}

Analogously to \eqref{eq:phi_Stratonovich_step}, the Stratonovich approach reads
\begin{equation} \label{eq:general_strato_step}
\begin{aligned}
    \text{(Stratonovich)} \qquad \Phi_i(N_+) = \Phi_i(N)
    &+ \frac{1}{2}\qty{\mu_i[\Phi_j(N),N] + \mu_i[\Phi_j(\Nplus),\Nplus]} \dd N \\
    &+ \frac{1}{2}\qty{\sigma_i^{\alpha}[\Phi_j(N),N] + \sigma_i^{\alpha}[\Phi_j(\Nplus),\Nplus]}\sqrt{\dd N}\xi_{\alpha,N} \, . \phantom{\text{(Stratonovich)} \qquad}
\end{aligned}
\end{equation}
To compare \eqref{eq:general_strato_step} to \eqref{eq:general_ito_step}, we wish to expand the $\Nplus$ functions, keeping only contributions up to order $\dd N$, analogously to \eqref{eq:phi_Stratonovich_itoed_step}. The expansions read
\begin{subequations}\label{eq:genral_expansion}
\begin{align}
\label{eq:mu_expansion}
\begin{split}
    \mu_i[\Phi_j(\Nplus),\Nplus] &= \mu_i[\Phi_j(N),N]
    + \frac{\partial}{\partial N}\mu_i[\Phi_j(N),N]\times\dd N \\
    &\phantom{=} + \frac{\partial}{\partial \Phi_l}\mu_i[\Phi_j(N),N] \times\dd\Phi_l + \mathcal{O}(\dd N^2, \dd N \dd \Phi_j, \dd \Phi_j^2) \, ,
\end{split} \\
\label{eq:sigma_expansion}
\begin{split}
    \sigma_i^\alpha[\Phi_j(\Nplus),\Nplus] &= \sigma_i^\alpha[\Phi_j(N),N]
    + \frac{\partial}{\partial N}\sigma_i^\alpha[\Phi_j(N),N]\times\dd N \\
    &\phantom{=}+ \frac{\partial}{\partial \Phi_l}\sigma_i^\alpha[\Phi_j(N),N] \times\dd\Phi_l + \mathcal{O}(\dd N^2, \dd N \dd \Phi_j, \dd \Phi_j^2) \, ,
\end{split} \\
    \dd \Phi_i \equiv \Phi_i(\Nplus) \, - \, &\Phi_i(N) \, .
\end{align}
\end{subequations}
Since the $\mu_i$ contributions in \eqref{eq:general_strato_step} already scale as $\dd N$, we only need to keep the leading term in the $\mu_i$ expansion \eqref{eq:mu_expansion}, so these contributions agree trivially between the \Ito\ and Stratonovich formulas, just as in the Markovian case.
As in the Markovian case, we need to be careful with the $\sigma_i^\alpha$ contributions. Plugging the leading term of \eqref{eq:sigma_expansion} into the Stratonovich formula \eqref{eq:general_strato_step} again matches the \Ito\ case \eqref{eq:general_ito_step}, but the additional terms in \eqref{eq:sigma_expansion} could give additional $\mathcal{O}(\sqrt{\dd N})$ contributions that combine with the overall $\sqrt{\dd N}$ factor to give a new drift contribution, like in \eqref{eq:phi_Stratonovich_itoed_step}. Clearly, the only possible $\mathcal{O}(\sqrt{\dd N})$ contribution comes from the third term in the $\sigma_i^\alpha$ expansion, from $\dd \Phi_l$. Expanding $\dd \Phi_l$ iteratively with \eqref{eq:general_strato_step}, this leading contribution is
\begin{equation} \label{eq:sigma_expansion_term}
    \sigma_i^\alpha[\Phi_j(\Nplus),\Nplus] \sim
    \frac{\partial}{\partial \Phi_l}\sigma_i^\alpha[\Phi_j(N),N] \times\dd\Phi_l \sim
    \frac{\partial}{\partial \Phi_l}\sigma_i^\alpha[\Phi_j(N),N] \times
    \sigma_l^\alpha[\Phi_j(N),N]\sqrt{\dd N}\xi_{\alpha,N} \, .
\end{equation}
Crucially, $\sigma_i^\alpha$ only depends on $\delta\phi_R$ and $\delta\pi_R$ (the current perturbation values) as per \eqref{eq:general_markovian_form_unpacked}, so the $\Phi_l$ derivative is only non-zero for $l=\delta\phi_R, \delta\pi_R$. However, by \eqref{eq:general_markovian_form_unpacked}, $\sigma_l^\alpha$ is zero for these $l$ values (the perturbations evolve classically without noise). Thus \eqref{eq:sigma_expansion_term} vanishes for all $i$, and no extra drift terms emerge.

Up to $\mathcal{O}(\dd N)$, the Stratonovich step \eqref{eq:general_strato_step} then reduces to the \Ito\ result \eqref{eq:general_ito_step}. For the same noise realizations $\xi_{\alpha,N}$, both approaches give equivalent evolutions for the background and the mode functions in the $\dd N \to 0$ limit. This makes sense physically: the stochastic noise depends on the full time evolution of the mode functions $\delta\phi_k$ from vacuum to coarse graining, and the noise at the last time step only has a minimal (subleading order) effect on the mode's final value. The connection between the noise and the current state of the system is less direct than in the Markovian approximation, which exaggerated the importance of the final kick by making the noise depend directly on $\phi_R$. It does not matter whether $\delta\phi_k$ is evaluated at $N$ or $\Nplus$, at the previous or later $\phi_R$, so to speak, and the \Ito\ and Stratonovich approaches (and all others following a similar pattern) become equal.

Finally, the alternating zoom-in scheme can be written as
\begin{equation} \label{eq:general_alternating_step}
\begin{aligned}
    \text{(Alternating)} \qquad \tilde{\Phi}_i(\Nplus) &= \Phi_i(N) + \mu_i[\Phi_i(N),N]\dd N \, , \\
    \Phi(\Nplus) &=  \tilde{\Phi}_i(\Nplus) + \sigma_i^\alpha[\tilde{\Phi}_j(\Nplus),N]\sqrt{\dd N} \, \xi_{\alpha,N} \, , \phantom{\qquad\text{(Alternating)}}
\end{aligned}
\end{equation}
analogously to \eqref{eq:phi_alternating_step}. A simple expansion again yields \eqref{eq:general_ito_step}, confirming the alternating scheme is also equal to the \Ito\ one.

One interpretation of these results is that the extra drift in the Markovian Stratonovich approach \eqref{eq:phi_Stratonovich_itoed_step} is an artifact of the Markovian approximation, absent in a more complete analysis of the system. This is why I advocate for the \Ito\ interpretation instead of the Stratonovich one in the Markovian case: it is better motivated by an analysis starting from the full system \eqref{eq:Friedmann_stochastic}, and it is equivalent to the unambiguously defined alternating zoom-in scheme also in the Markovian limit.

Another interpretation is that since the \Ito\ and Stratonovich approaches are equal in the full non-Markovian case, they should also be equal in the Markovian case. Cases where this is not true---for example, the super-Planckian dynamics of secion~\ref{sec:how_big_difference}---must break the Markovian approximation, as discussed in Section~\ref{sec:validity_of_markovian}, and the full non-Markovian equations should be used instead.
In slow-roll models, the approaches only become different when $H$ changes quickly, and this is exactly the limit where the slow-roll approximation for the mode functions \eqref{eq:sr_mode_functions} breaks down.
The difference between \Ito\ and Stratonovich is then beyond the accuracy of the Markovian approximation, as suggested in \cite{Winitzki:1995pg, Vennin:2015hra}. When the Markovian approximation holds and the approaches are equal, it is fine to use the Stratonovich one, if this is technically simpler, e.g., to conserve the chain rule \cite{Pinol:2018euk, Pinol:2020cdp}.

Since we only used the general form of the equations \eqref{eq:general_markovian_form}, \eqref{eq:general_markovian_form_unpacked}, these results are general, and apply, e.g., for non-canonical inflation models and different window functions. The generalization to multifield inflation and spectator fields is also trivial.

\section{Discussion and conclusions}
\label{sec:discussion}

In this paper, I have explored the origin of the stochastic noise in stochastic inflation. The noise arises from the time-dependent coarse-graining scale, which continuously lets short-wavelength modes join the coarse-grained background. I have considered this process systematically, introducing a zoom-in scheme that alternates between the system's classical evolution, where the comoving coarse-graining scale stays fixed, and zoom-in steps, where the coarse-graining scale changes instantaneously.
This conceptual development is a step towards a systematic derivation of the stochastic inflation formalism from first principles. It also guides the numerical implementation of stochastic inflation---indeed, this description was used in the numerical work of \cite{Figueroa:2020jkf, Figueroa:2021zah}.

I also discussed the competing \Ito\ and Stratonovich interpretations of the stochastic equations. I explained the difference between the interpretations in depth in Appendix~\ref{sec:stochastic_formulas}.
In the main text, I applied the interpretations to stochastic single-field inflation in the Markovian slow-roll limit, showing that the alternating zoom-in scheme matches the \Ito\ interpretation, while the Stratonovich interpretation does not seem to have a similar match.
On the other hand, I demonstrated with both an analytical computation and numerical simulations that the \Ito\ and Stratonovich interpretations only differ significantly at super-Planckian energies, where the standard slow-roll form of the stochastic equations is, in any case, expected to become inaccurate. Similar arguments about the inaccuracy were presented earlier in \cite{Salopek:1990re, Winitzki:1995pg, Vennin:2015hra, Pinol:2018euk}.

I then considered the general non-Markovian case, where the short-wavelength modes evolve in the stochastic background, allowing backreaction between the two.
To use the Markovian framework with the full equations, the modes must be promoted to equal status with the background quantities.
The stochastic noise is then only indirectly coupled to the background, and I showed the \Ito\ and Stratonovich interpretations are equal. The alternating zoom-in scheme successfully describes both. The same description applies also for multifield setups.

Before concluding, let me briefly compare this work to earlier literature.

\paragraph{Previous literature on \Ito\ versus Stratonovich.}
The difference between the \Ito\ and Stratonovich interpretations of stochastic inflation has been discussed in a handful of papers over the years.
In the early work \cite{Salopek:1990re}, the authors advocated for the \Ito\ approach, 
arguing it better matches their formulation of stochastic inflation which was similar to this work, see the discussion in Section~\ref{sec:alternating_scheme_introduced}. In \cite{Mezhlumian:1991hw}, it was argued that the Stratonovich approach is the more natural one since the white noise approximation is the limit of a more general colored process, which the Stratonovich process mimics. However, as explained in \cite{Kampen:1981}, such limits depend strongly on the way they are taken, and the connection between colored noise and the Stratonovich process is not a straightforward one.
The Stratonovich approach was also adopted in \cite{Linde:1993xx},
while \cite{Garriga:1997ef, Vilenkin:1999kd, Winitzki:2008zz} opted for \Ito, finding certain eternal-inflation observables to be independent of the time parametrization in this approach.
The authors of \cite{Fujita:2014tja, Tokuda:2017fdh} advocated for the \Ito\ approach based on a causality argument similar (if less detailed) to the approach of the current paper,
while in \cite{Pinol:2018euk, Pinol:2020cdp}, it was argued that the Stratonovich approach is needed to preserve the field-space covariance of the stochastic equations.

The previous arguments were based on heuristics and consistency checks.
In this paper, I have given a first-principles interpretation for the \Ito\ approach as the alternating zoom-in scheme. I have also emphasized that the stochastic framework is only an approximation of the underlying dynamics. It may be inconsistent with some properties of the full, fundamental theory; the goal is simply to minimize such errors while carefully tracking the coarse-graining procedure. As noted above, the differences between the \Ito\ and Stratonovich approaches vanish in the full non-Markovian case.

\paragraph{Fictionality of stochastic velocities.}
The zoom-in scheme presented in this paper highlights the correct physical interpretation of the stochastic noise in stochastic inflation: the noise describes a change in perspective, a `zooming in' into an expanding patch of space, and it is a priori independent of the inflaton's classical time evolution. In \cite{Cohen:2021jbo, Cohen:2022clv}, it was claimed that if the field fluctuations are particularly strong, they induce a high stochastic field velocity, and the related energy may exceed the cutoff scale of the effective field theory and thus invalidate the stochastic approach.
The problem with this argument is that there is no physical energy scale related to the stochastic noise.
The processes that produce the noise operate in accordance with usual cosmological perturbation theory: high-energy perturbations are in their vacuum state deep inside the Hubble radius, and they only start to grow around the Hubble exit, when their energy has been diluted by the expansion of space.
The induced---possibly large---time derivative of the coarse-grained field is due to the change in the coarse-graining scale, which we can choose arbitrarily at any given time, following our chosen zoom-in scheme. No physical energy scale is introduced by this choice---indeed, if it were, our alternating zoom-in scheme would instantly lead to an infinite energy when the coarse-grained field changes in zero time during the zoom-in step. Fast stochastic evolution alone can't invalidate stochastic inflation. At most, strong perturbations may lead to large \emph{spatial} derivatives, causing problems with the separate universe approach \cite{Jackson:2023obv}, but this problem vanishes for a coarse-graining scale far enough outside the Hubble radius. 
Extremely strong gradients could end inflation locally \cite{Linde:1993xx}, providing a natural cutoff mechanism, but in practice, typical stochastic deviations tend to arise from gradual change over long scales rather than an abrupt jump \cite{Tomberg:2022mkt}.
A high stochastic velocity is also not automatically inherited by the classical field if stochasticity is turned off, e.g., by the end of inflation.

The setup resembles the famous lighthouse paradox (see, e.g., \cite{Zakamska:2015nys}): if a rotating lighthouse shines a beam of light on a distant shore, the beam's end may move at a superluminal speed\footnote{A more realistic version of the paradox replaces the lighthouse with a laser and the shore with the Moon.}. However, the photons themselves don't travel faster than light; instead, different photons arrive at different end points at almost the same time. The `beam's end' is an illusion, useful for describing the workings of a lighthouse, but it follows different laws than individual photons. Similarly, the value of the stochastic inflaton field is an illusion, set by the human-made decision of a coarse-graining scale.

The zoom-in procedure resembles the renormalization group calculations of quantum field theory (see, e.g., \cite{Peskin:1995ev}).
In both cases, high-energy degrees of freedom are integrated over to obtain an effective theory for the low-energy ones.
The two halves of the system are separated by a bookkeeping variable---the renormalization or coarse-graining scale---without inherent physical meaning.
Changing the renormalization scale doesn't change the predictions of the theory, but it may make a perturbation theory expansion converge faster. A result computed to a fixed loop order will be more accurate if the renormalization scale is chosen properly. Similarly, stochastic inflation produces more accurate results if the coarse-graining scale (zoom-in scheme) is chosen appropriately.
It would be interesting to study this link in more detail.

\paragraph{Different zoom-in schemes.}
One may ask if there are other reasonable zoom-in schemes besides the alternating one described in this paper.
If the alternating nature of the scheme is abandoned completely, interpreting the noise becomes difficult. However, the lengths of the classical steps and the value of the coarse-graining scale after each zoom-in step could be varied. This could improve the accuracy of the approximation and the numerical convergence in the $\dd N \to 0$ limit, especially if the Hubble parameter $H$ varies a lot during the computation.

\begin{figure}
    \centering
    \includegraphics{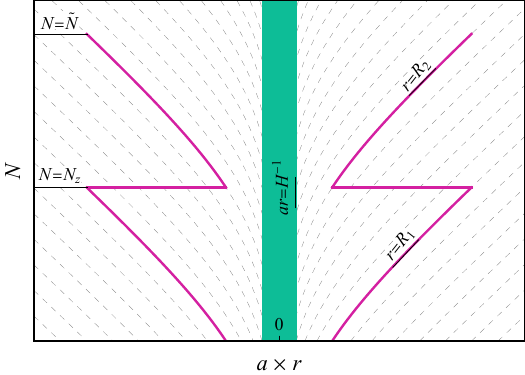}
    \caption{The coarse-graining scale $R$ versus the time $N$ in a two-step zoom-in scheme, analogously to Figure~\ref{fig:zoom}.}
    \label{fig:zoom_2}
\end{figure}

Instead of taking the $\dd N \to 0$ limit, one may imagine leaving the length of the classical evolution steps finite.
In particular, we may divide the full inflationary period into two periods of classical evolution separated by a single zoom-in, as depicted in Figure~\ref{fig:zoom_2}. During the first span of classical evolution, we fix the comoving size of the inflating patch to $R_1$, and evolve linear modes with wave numbers between $k_{R_1}$ and $k_{R_2}$ to super-Hubble scales, for some $R_2 < R_1$. In the zoom-in step, at time $N_z$, we then change the coarse-graining scale from $R_1$ to $R_2$, integrating in all the modes between $k_{R_1}$ and $k_{R_2}$ in one macroscopic jump. We end by evolving classically, to obtain quantities coarse-grained at scale $R_2$.
This is a middle ground between linear perturbation theory and full stochastic inflation, see Figure~\ref{fig:zoom}. In fact, this is essentially what is done in the `classical $\Delta N$ formalism,' see, e.g., \cite{Cai:2018dkf, Atal:2019cdz, Atal:2019erb, Biagetti:2021eep, Hooshangi:2021ubn, Hooshangi:2022lao, Pi:2022ysn, Firouzjahi:2023ahg, Hooshangi:2023kss, Pi:2024jwt, Artigas:2024xhc, Inui:2024sce, Ballesteros:2024pwn}, where an initial Gaussian field distribution is evolved classically from an initial time (our $N=N_z$) until the end of inflation. In these computations, the width of the Gaussian is usually treated as a free input parameter. The current description tells us which linear modes
should make up the initial distribution. In particular, only modes that are super-Hubble at the initial time $N_z$ can contribute to the distribution's width, and the initial Hubble scale gives the shortest accessible coarse-graining scale.
This description could also be generalized to multiple (still finite) cycles of classical evolution and macroscopic jumps.
Such a system would be simpler to solve than full stochastic inflation, maybe even analytically, and it might still provide a reasonable description of the physics if the cycles were chosen optimally. I leave such considerations for future work.

\paragraph{Deriving the stochastic formalism from first principles.}
Finally, the concept of a zoom-in scheme can be useful when deriving stochastic inflation from first principles.
A full derivation would shed light on the accuracy of stochastic computations and the limitations of the separate universe and white noise approximations.
Many derivations have been presented in the literature, using varying levels of approximation, see, e.g., \cite{Tokuda:2018eqs, Cruces:2021iwq, Pinol:2020cdp, Launay:2024qsm} for recent papers. The nature of the stochastic noise often receives little attention in such works. This paper aims to amend the issue, paving the way for a deeper understanding of the relationship between stochastic inflation and fundamental theories.

\acknowledgments

This work was supported by the Lancaster--Manchester--Sheffield Consortium for Fundamental Physics under STFC grant: ST/T001038/1.

\appendix

\section{Deriving results for stochastic differential equations}
\label{sec:stochastic_formulas}

In this appendix, I review basic results from stochastic calculus. For more details, see, e.g., \cite{Ito:1944, Stratonovich:1966, Lawler:2014, Martingales:2023}.

Let us study the following generic SDE:
\begin{equation} \label{eq:generic_SDE}
    \dd X = \mu(X,t)\dd t + \sigma(X,t)\dd W_t \, .
\end{equation}
This is the increment in the stochastic variable $X$ in a time step $\dd t$. The first term gives a deterministic drift, while the second term is stochastic: $\dd W_t$ is the time step of a \emph{Wiener process} \cite{Lawler:2014}. The steps are independent Gaussian random variables with
\begin{equation}
    \dd W_t = \sqrt{\dd t} \, \xi_t \, , 
    \qquad \expval{\xi_t} = 0 \, ,
    \qquad \expval{\xi_t \xi_{t'}} = \delta_{tt'} \, .
\end{equation}
It follows that
\begin{equation} \label{eq:mean_var_dx}
    \expval{\dd X} = \mu \, \dd t \, , \qquad \text{Var} \, X = \sigma^2 \dd t \, ,
\end{equation}
that is, $\mu$ is the mean time increment and $\sigma^2$ is its variance, when multiplied by $\dd t$.

To understand the $\dd t$ scalings in \eqref{eq:mean_var_dx}, consider a finite time interval $\Delta t$ small enough that $\mu$ and $\sigma$ are approximately constant within it, and divide it into $n$ sub-intervals of length $\dd t = \Delta t/n$. The integral of \eqref{eq:generic_SDE} over the interval gives
\begin{equation} \label{eq:longer_delta_t}
    \Delta X
    = \mu \Delta t + \sigma \sqrt{\Delta t} \, \bar{\xi} \, ,
    \qquad
    \bar{\xi} \equiv \frac{1}{\sqrt{n}}\sum_{i=1}^{n} \xi_i \, .
\end{equation}
Since $\bar{\xi}$ is a sum of independent, zero-mean Gaussian random variables, it is itself a zero-mean Gaussian random variable, with variance equal to the sum of the components' variances. Hence,
\begin{equation} \label{eq:longer_var_x}
    \text{Var} \, \bar{\xi} = \frac{1}{n}\times n \times 1 = 1 \, ,
\end{equation}
and the left equation in \eqref{eq:longer_delta_t} is of the same form as \eqref{eq:generic_SDE} and independent of $n$. In the continuum limit of small time intervals, the finite-time behavior of $X$ converges to a well-defined form, thanks to the scaling of the variance in \eqref{eq:mean_var_dx}.

The different scalings of the two terms in \eqref{eq:generic_SDE} may seem baffling: the first scales as $\dd t$, while the second scales as $\sqrt{\dd t}$. The first dominates in the long-time limit $\dd t \to \infty$, while the second dominates in the continuum limit $\dd t \to 0$. Note, however, that the terms contribute quite differently: the first one provides the mean behavior of $X$, and the second gives the variation around this mean. Both are needed to characterize the stochastic process. This is an important rule of thumb below, where I derive some central results of stochastic calculus at a `physicist's accuracy': \emph{when expanding in $\dd t$, one must keep terms of both orders, $\sqrt{\dd t}$ and $\dd t$}, to catch both the mean and the variations. Higher-order terms can be neglected as subleading corrections.

Another way to see this is to note that even powers of the noises $\xi_i$ yield deterministic quantities in the $\dd t \to 0$ limit. To see this, let us perform a trick: replace the Gaussian $\xi_t$ with binary variables $\hat{\xi}_t$ with the same mean and variance, that is,
\begin{equation} \label{eq:xi_hat}
    \hat{\xi}_t \in \qty{-1, 1} \, ,
    \quad p(\hat{\xi}_t = 1) = p(\hat{\xi}_t = -1) = \frac{1}{2} \,
    \quad \Rightarrow
    \quad \expval{\xi_t} = 0 \, ,
    \quad \expval{\xi_t \xi_{t'}} = \delta_{tt'} \, .
\end{equation}
The central limit theorem tells us that the sum of many $\hat{\xi}_i$ variables is still a Gaussian, with a variance equal to the sum of the component variances. For finite time intervals, the binary components merge into a combined Gaussian just the same as the Gaussian components did in \eqref{eq:longer_delta_t}. We straightforwardly see that
\begin{equation} \label{eq:xi_powers}
    \hat{\xi}_t^n = 
    \begin{cases}
        1, &\text{$n$ even,}\\
        \hat{\xi}_i, &\text{$n$ odd,}
    \end{cases}
\end{equation}
and we can use the same result for $\xi_t^n$ in the $\dd t \to 0$ limit. Any $\dd t$ expansion starting from \eqref{eq:generic_SDE} thus naturally produces only two types of terms: deterministic ones, with a leading behavior $\propto \dd t$, and stochastic ones proportional to $\xi_i$, with a leading behavior $\propto \sqrt{\dd t}$. We will see this principle in action in the next section.

To solve equation \eqref{eq:generic_SDE} between two times $t_1$ and $t_2$ in a case where $\mu$ and $\sigma$ are not constants, we need to iterate over it, solving $X$ one-time step at a time. We start with an initial value $X(t_1)$ at $t_1$, divide the time interval into small steps of length $\dd t$, and compute $X(t+\dd t)$ by plugging in $t$ and $X$ evaluated (or estimated) at some point in the interval $[t, t+\dd t]$ to the right-hand side of \eqref{eq:generic_SDE} to obtain $X(t+\dd t) = X(t) + \dd X$. However, there is an ambiguity here: at which point in the interval should the right-hand side of \eqref{eq:generic_SDE} be evaluated? For usual differential equations, solved by Riemann integration, this does not matter: any choice converges to the same final answer when $\dd t \to 0$. The final result is the same whether we choose to approximate the integrand by its value at the start or the end of an infinitesimal interval or by some linear combination of these. However, this is not true for stochastic variables.

\subsection{\Ito\ versus Stratonovich}
Let us introduce two integration schemes for \eqref{eq:generic_SDE}. In the \emph{\Ito} approach, we evaluate $X$ and $t$ at the beginning of the $\dd t$ interval, leading to the Euler method. In other words,
\begin{equation} \label{eq:Ito_step}
    \text{(\Ito)} \qquad X(\tplus) = X(t) + \mu[X(t),t]\dd t + \sigma[X(t),t]\sqrt{\dd t} \, \xi_t \, , \quad
    \tplus \equiv t + \dd t \, . \phantom{\qquad \text{(\Ito)}}
\end{equation}
This is simple: to know the value of $X$ in the future, we only need to know the value of $X$ now.

In the \emph{Stratonovich} approach, we instead use the averages of $\mu$ and $\sigma$ evaluated at the start and end of the $\dd t$ interval \cite{Stratonovich:1966}:
\begin{equation} \label{eq:Stratonovich_step_estimator}
\begin{aligned}
    X(\tplus) = X(t)
    &+
    \frac{1}{2}\qty{\mu[X(t),t]+\mu[X(\tplus),\tplus]}\dd t \\
    &+
    \frac{1}{2}\qty{\sigma[X(t),t]+\sigma[X(\tplus),\tplus]}\sqrt{\dd t} \, \xi_t \, .
\end{aligned}
\end{equation}
For a pictorial comparison of \eqref{eq:Ito_step} and \eqref{eq:Stratonovich_step_estimator}, see Figure~\ref{fig:schema_integral_bin}.

Equation \eqref{eq:Stratonovich_step_estimator} is somewhat convoluted since it seems we're required to already know the value of $X$ in the future to determine $X$ there. However, we can solve the equation iteratively. The first iteration is enough to get the correct continuum limit, and it yields the Euler--Heun method, where $X(\tplus)$ is replaced by the Euler estimate \eqref{eq:Ito_step} on the right-hand side.
We can then expand the right-hand side of \eqref{eq:Stratonovich_step_estimator} to the leading order in $\dd t$ to obtain
\begin{equation} \label{eq:Stratonovich_step_leading}
    X(\tplus) = X(t) + \mu\,\dd t +
    \frac{\sigma}{2}\frac{\partial\sigma}{\partial X}\xi_t^2\,\dd t + \sigma\sqrt{\dd t} \, \xi_t \, .
\end{equation}
To lighten the notation, I will drop the explicit arguments of $\mu$ and $\sigma$ from now on---unless otherwise noted, these are to be evaluated at $[X(t),t]$. As stressed above, the `leading order in $\dd t$' includes terms of order $\sqrt{\dd t}$.

The new term $\propto \xi_t^2$ arises from the mixing of the last terms in \eqref{eq:Ito_step} and \eqref{eq:Stratonovich_step_estimator}. As we saw in the last section, we can replace $\xi_t^2$ by 1: the new term is deterministic and contributes to the mean behavior of $X$. This gives us the final form
\begin{equation} \label{eq:Stratonovich_step_clean}
    \text{(Stratonovich)} \qquad X(\tplus) = X(t) + \qty(\mu + \frac{\sigma}{2}\frac{\partial\sigma}{\partial X})\dd t
    + \sigma\sqrt{\dd t} \, \xi_t \, . \phantom{\text{(Stratonovich)} \qquad}
\end{equation}
In \eqref{eq:Stratonovich_step_clean}, we have converted \eqref{eq:Stratonovich_step_estimator} into an Euler-like form, where $X(\tplus)$ only depends on $X(t)$ and $t$. This resembles the \Ito\ solution \eqref{eq:Ito_step}. However, the non-trivial behavior of the noise terms has introduced an additional contribution to the drift. The original equation \eqref{eq:generic_SDE} is not well-defined until either the \Ito\ or the Stratonovich approach has been chosen; different approaches clearly give different time evolutions and correspond to different physical systems. On the other hand, an equation defined in one approach can always be converted to an equation in the other by adding or subtracting the additional drift term.

Taking different linear combinations of $\mu$ and $\sigma$ evaluated at $t$ and $\tplus$ creates additional approaches on top of \Ito\ and Stratonovich. However, these two are the most common ones: \Ito\ because of the simple correspondence between the `base equation' \eqref{eq:generic_SDE} and the time evolution \eqref{eq:Ito_step}, and Stratonovich because of an issue related to the chain rule, which I will tackle next.

\subsection{The chain rule for stochastic variables}
Let us define a new stochastic variable $Y(t) \equiv f[X(t),t]$, where $f$ is a smooth function. According to the chain rule of calculus, the time steps of $Y$ should be given by
\begin{equation} \label{eq:chain_rule}
    \dd Y = \frac{\partial f}{\partial t} \dd t + \frac{\partial f}{\partial X} \dd X \, .
\end{equation}
However, expanding $\dd Y$ to leading order in $\dd t$ with the \Ito\ approach \eqref{eq:Ito_step} gives\footnote{Like $\mu$ and $\sigma$, $f$ is to be evaluated at $\qty[X(t),t]$ from now on.}
\begin{equation} \label{eq:Itos_lemma}
\begin{aligned}
    Y(\tplus) - Y(t) &=
    \frac{\partial f}{\partial t} \,\dd t +
    \frac{\partial f}{\partial X} \,\dd X + \frac{1}{2}\frac{\partial^2 f}{\partial X^2} \qty(\dd X)^2 + \dots \\
    &= \qty(\frac{\partial f}{\partial t} + \mu\frac{\partial f}{\partial X} + \frac{\sigma^2}{2}\frac{\partial^2 f}{\partial X^2}) \dd t
    + \sigma\frac{\partial f}{\partial X} \sqrt{\dd t}\,\xi_t
    + \mathcal{O}(\dd t^{3/2}) \, .
\end{aligned}
\end{equation}
The noise terms strike again: we had to include the $(\dd X)^2$ term to capture the extra $(\sqrt{\dd t}\,\xi_t)^2$ contribution to the drift. (We again used $\xi_t^2 \to 1$.) This ruins the chain rule \eqref{eq:chain_rule}. Result \eqref{eq:Itos_lemma} is known as \emph{\Ito's lemma} \cite{Ito:1944}.

What about the Stratonovich approach? Let us first introduce a bit of notation:
\begin{equation} \label{eq:Stratonovich_circle}
    g \overset{X}{\circ} \dd W_t \equiv \frac{g}{2}\frac{\partial g}{\partial X} \dd t + g\,\dd W_t
\end{equation}
for a generic function $g[X(t),t]$. With this, the Stratonovich step \eqref{eq:Stratonovich_step_clean} can be written in the simple form
\begin{equation} \label{eq:Stratonovich_step_circled}
    \text{(Stratonovich)} \qquad X(\tplus) - X(t) = \mu \, \dd t
    + \sigma \overset{X}{\circ} \dd W_t \, , \phantom{\text{(Stratonovich)} \qquad}
\end{equation}
directly analogous to the base equation \eqref{eq:generic_SDE}. Using \eqref{eq:Stratonovich_step_clean} and \eqref{eq:Stratonovich_step_circled}, the chain rule takes the form\footnote{Except for explicit $\frac{\partial}{\partial t}$ terms, the time $t$ is kept constant in all partial derivatives in \eqref{eq:Stratonovich_chain_rule}.}
\begin{equation} \label{eq:Stratonovich_chain_rule}
\begin{aligned}
    Y(\tplus) - Y(t) &=
    \frac{\partial f}{\partial t} \dd t +
    \frac{\partial f}{\partial X} \dd X + \frac{1}{2}\frac{\partial^2 f}{\partial X^2} \qty(\dd X)^2 + \dots \\
    &= \qty(\frac{\partial f}{\partial t} + \mu\frac{\partial f}{\partial X} + \frac{\sigma}{2}\frac{\partial \sigma}{\partial X}\frac{\partial f}{\partial X} + \frac{\sigma^2}{2}\frac{\partial^2 f}{\partial X^2}) \dd t
    + \sigma\frac{\partial f}{\partial X} \sqrt{\dd t}\,\xi_t
    + \mathcal{O}(\dd t^{3/2}) \\
    &= \qty(\frac{\partial f}{\partial t} + \mu\frac{\partial f}{\partial X}
    + \frac{\sigma}{2}\frac{\partial f}{\partial X}\frac{\partial}{\partial f}\qty[\sigma\frac{\partial f}{\partial X}]) \dd t
    + \sigma\frac{\partial f}{\partial X} \sqrt{\dd t}\,\xi_t
    + \mathcal{O}(\dd t^{3/2}) \\
    &= \frac{\partial f}{\partial t}\dd t +
    \mu\frac{\partial f}{\partial X}\dd t + \sigma \frac{\partial f}{\partial X} \overset{Y}{\circ} \dd W_t \, .
\end{aligned}
\end{equation}
We see that \eqref{eq:chain_rule} holds when the circle operator on the last line is taken with respect to $Y=f$ and operates on the whole combination $\sigma\partial_X f$ to its left.

As an interesting consistency check, let us consider the case where $\sigma$ is a constant, independent of $X$. Since $\partial_X\sigma = 0$, the two approaches \eqref{eq:Ito_step} and \eqref{eq:Stratonovich_step_clean} obviously coincide for $X$. Since $X$ and $Y$ are in a one-to-one relationship, the approaches should also be equivalent for $Y$. This is not a trivial statement, since the standard deviation of $\dd Y$, $\sigma_Y=\sigma\partial_X f$ from \eqref{eq:Stratonovich_chain_rule}, may depend on the variables. Nevertheless, we can easily see that the $Y$ evolutions coincide by comparing the second rows of \eqref{eq:Itos_lemma} and \eqref{eq:Stratonovich_chain_rule} and using $\partial_X\sigma = 0$. The extra term in \Ito's lemma \eqref{eq:Itos_lemma} is hidden in the Stratonovich differential on the last line of \eqref{eq:Stratonovich_chain_rule}.

\subsection{Alternating deterministic evolution and stochastic kicks}
The alternating scheme discussed elsewhere in this paper, that is, taking alternating steps with deterministic evolution and stochastic kicks, can be written as
\begin{equation} \label{eq:alternating_step}
\begin{aligned}
    \text{(Alternating)} \qquad &\tilde{X}(\tplus) = X(t) + \mu[X(t),t]\dd t \, , \phantom{\qquad \text{(Alternating)}} \\
    &X(\tplus) = \tilde{X}(\tplus) + \sigma[\tilde{X}(\tplus),\tplus]\sqrt{\dd t} \, \xi_t \, .
\end{aligned}
\end{equation}
We first added to $X(t)$ the $\mu$ term as the classical Euler time step and then used the thus obtained value $\tilde{X}$ as the $X$-position when taking the stochastic step. This resembles the Euler--Heun method discussed above for the Stratonovich approach, but the crucial difference is that $\tilde{X}$ does not contain a noise component with the power $\sqrt{\dd t}$. Hence, the leading order result for $X(\tplus)$ obtains no non-trivial corrections from the noise and is reduced to the \Ito\ formula \eqref{eq:Ito_step}. The alternating steps approach is equivalent to the \Ito\ approach.

\subsection{Fokker--Planck equation}

The statistics of the stochastic variable $X$ at time $t$ are described by its probability distribution $P(X,t)$. Let us consider the time evolution of this distribution. To get $P(X,\tplus)$, we consider all values $\tilde{X}$ at the previous time $t$ and all possible steps $\Delta X$ that lead from $\tilde{X}$ to $X$. The steps depend on the noise variable $\xi_t$. Integrating over the probability distributions of $\tilde{X}$ and $\xi_t$ while maintaining a fixed $X$ at $\tplus$ gives
\begin{equation} \label{eq:Fokker_Planck_time_step}
    P(X,\tplus) = \int \dd \tilde{X} \int \dd \xi_t \, P(\xi_t) P(\tilde{X}, t) \delta(\tilde{X} + \Delta X - X) \, .
\end{equation}
Here $P(\xi_t)=\frac{1}{\sqrt{2\pi}}e^{-\xi_t^2/2}$. The step $\Delta X$ is given by the Langevin equation; in the \Ito\ approach, \eqref{eq:Ito_step} gives
\begin{equation} \label{eq:Ito_step_for_FP}
    \Delta X = \mu(\tilde{X},t)\dd t + \sigma(\tilde{X},t)\sqrt{\dd t} \, \xi_t \, .
\end{equation}
Next, we expand the delta function in \eqref{eq:Fokker_Planck_time_step} in powers of $\dd t$ around $\tilde{X}=X$:
\begin{equation} \label{eq:delta_expansion}
\begin{aligned}
    \delta(\tilde{X}+\Delta X-X) &= \delta(\tilde{X}-X) + \delta'(\tilde{X}-X)\Delta X + \frac{1}{2}\delta''(\tilde{X}-X)\Delta X^2 + \dots \\
    &= \delta(\tilde{X}-X)
    + \delta'(\tilde{X}-X)\qty(\mu\,\dd t + \sigma\sqrt{\dd t} \, \xi_t) \\
    &\phantom{=} + \frac{1}{2}\delta''(\tilde{X}-X)\sigma^2 \xi_t^2 \dd t + \mathcal{O}(\dd t^{3/2}) \, .
\end{aligned}
\end{equation}
Integrating over $\xi_t$, the term linear in $\xi_t$ vanishes and the quadratic term gives $\int \dd \xi_t P(\xi_t) \xi_t^2 = 1$. Then, in the small time step limit, \eqref{eq:Fokker_Planck_time_step} becomes
\begin{equation} \label{eq:Fokker_Planck_time_step_2}
\begin{aligned}
    P(X,\tplus) &= \int \dd \tilde{X} \, P(\tilde{X}, t) \delta(\tilde{X} - X) \\
    &\phantom{=} + \int \dd \tilde{X} \, P(\tilde{X}, t) \delta'(\tilde{X} - X)\mu(\tilde{X},t) \, \dd t \\
    &\phantom{=} + \frac{1}{2}\int \dd \tilde{X} \, P(\tilde{X}, t) \delta''(\tilde{X} - X)\sigma^2(\tilde{X},t) \, \dd t \\
    &= P(X,t) - \partial_X\qty[P(X,t)\mu(X,t)] \dd t + \frac{1}{2}\partial^2_X\qty[P(X,t)\sigma^2(X,t)] \dd t \, ,
\end{aligned}
\end{equation}
where I moved the derivatives from the delta functions to the integrands through partial integration. Moving the terms around, we get the Fokker--Planck equation \cite{Lawler:2014}
\begin{equation} \label{eq:Fokker_Planck_Ito}
    \partial_t P = \partial_X\qty[\frac{1}{2}\partial_X(\sigma^2P) - \mu P] \, . 
\end{equation}
In the Stratonovich approach, equation \eqref{eq:Ito_step_for_FP} is replaced by 
\begin{equation} \label{eq:Stratonovich_step_estimator_B}
    \Delta X = \mu(\tilde{X},t)\dd t +
    \frac{\sigma(\tilde{X},t)}{2}\frac{\partial\sigma(\tilde{X},t)}{\partial \tilde{X}}\xi_t^2\,\dd t + \sigma(\tilde{X},t)\sqrt{\dd t} \, \xi_t \, ,
\end{equation}
from \eqref{eq:Stratonovich_step_leading}, so that, in practice, the Stratonovich form of the Fokker--Planck equation is again obtained by replacing $\mu \to \mu + \frac{\sigma}{2}\frac{\partial \sigma}{\partial X}$ in \eqref{eq:Fokker_Planck_Ito}, giving\footnote{In \eqref{eq:Stratonovich_step_leading}, I replaced $\xi_t^2$ by $1$ by appealing to the nature of the noise; here, the same is straightforwardly accomplished by the $\xi_t$ integration.}
\begin{equation} \label{eq:Fokker_Planck_Stratonovich}
    \partial_t P = \partial_X\qty[\frac{\sigma}{2}\partial_X(\sigma P) - \mu P] \, . 
\end{equation}
This can easily be seen to be covariant under changes of the stochastic variable: with the above definition $Y(t) \equiv f[X(t),t]$, we have
\begin{equation} \label{eq:P_X_Y_transform}
    P_X = (\partial_X Y) P_Y \, , \quad
    \partial_X = (\partial_X Y)\partial_Y
\end{equation}
at a fixed time $t$, given by the general rules of differentiation and probability densities, where I labeled the quantities with $X$ or $Y$ depending on which variable they apply to. If the field transformation behaves expectedly, we also have
\begin{equation} \label{eq:X_Y_covariant_transform}
    \dd Y = (\partial_t Y) \dd t + (\partial_X Y) \dd X
    = \underbrace{\qty[(\partial_X Y)\mu_X + \partial_t Y]}_{\equiv \mu_Y} \dd t + \underbrace{(\partial_X Y)\sigma_X}_{\equiv \sigma_Y} \sqrt{\dd t}\,\xi_t \, .
\end{equation}
Plugging these into the Stratonovich equation \eqref{eq:Fokker_Planck_Stratonovich} with $X$-labels and rearranging terms produces an equation of the same form but with $Y$-labels. The same is not true for the \Ito\ version \eqref{eq:Fokker_Planck_Ito}. This is a manifestation of the covariance of the Stratonovich approach discussed above: \eqref{eq:X_Y_covariant_transform} holds there, but not in the \Ito\ approach.

\bibliographystyle{JHEP}
\bibliography{stoch}

\providecommand{\href}[2]{#2}\begingroup\raggedright\begin{thebibliography}{10}

\bibitem{Planck:2018jri}
{\scshape Planck} collaboration, \emph{{Planck 2018 results. X. Constraints on inflation}}, \href{https://doi.org/10.1051/0004-6361/201833887}{\emph{Astron. Astrophys.} {\bfseries 641} (2020) A10} [\href{https://arxiv.org/abs/1807.06211}{{\ttfamily 1807.06211}}].

\bibitem{Salopek:1990jq}
D.S.~Salopek and J.R.~Bond, \emph{{Nonlinear evolution of long wavelength metric fluctuations in inflationary models}}, \href{https://doi.org/10.1103/PhysRevD.42.3936}{\emph{Phys. Rev. D} {\bfseries 42} (1990) 3936}.

\bibitem{Wands:2000dp}
D.~Wands, K.A.~Malik, D.H.~Lyth and A.R.~Liddle, \emph{{A New approach to the evolution of cosmological perturbations on large scales}}, \href{https://doi.org/10.1103/PhysRevD.62.043527}{\emph{Phys. Rev. D} {\bfseries 62} (2000) 043527} [\href{https://arxiv.org/abs/astro-ph/0003278}{{\ttfamily astro-ph/0003278}}].

\bibitem{Sasaki:1995aw}
M.~Sasaki and E.D.~Stewart, \emph{{A General analytic formula for the spectral index of the density perturbations produced during inflation}}, \href{https://doi.org/10.1143/PTP.95.71}{\emph{Prog. Theor. Phys.} {\bfseries 95} (1996) 71} [\href{https://arxiv.org/abs/astro-ph/9507001}{{\ttfamily astro-ph/9507001}}].

\bibitem{Sasaki:1998ug}
M.~Sasaki and T.~Tanaka, \emph{{Superhorizon scale dynamics of multiscalar inflation}}, \href{https://doi.org/10.1143/PTP.99.763}{\emph{Prog. Theor. Phys.} {\bfseries 99} (1998) 763} [\href{https://arxiv.org/abs/gr-qc/9801017}{{\ttfamily gr-qc/9801017}}].

\bibitem{Lyth:2004gb}
D.H.~Lyth, K.A.~Malik and M.~Sasaki, \emph{{A General proof of the conservation of the curvature perturbation}}, \href{https://doi.org/10.1088/1475-7516/2005/05/004}{\emph{JCAP} {\bfseries 05} (2005) 004} [\href{https://arxiv.org/abs/astro-ph/0411220}{{\ttfamily astro-ph/0411220}}].

\bibitem{Starobinsky:1986fx}
A.A.~Starobinsky, \emph{{STOCHASTIC DE SITTER (INFLATIONARY) STAGE IN THE EARLY UNIVERSE}}, \href{https://doi.org/10.1007/3-540-16452-9_6}{\emph{Lect. Notes Phys.} {\bfseries 246} (1986) 107}.

\bibitem{Lawler:2014}
G.F.~Lawler, \emph{Stochastic calculus: An introduction with applications}, .

\bibitem{Fujita:2013cna}
T.~Fujita, M.~Kawasaki, Y.~Tada and T.~Takesako, \emph{{A new algorithm for calculating the curvature perturbations in stochastic inflation}}, \href{https://doi.org/10.1088/1475-7516/2013/12/036}{\emph{JCAP} {\bfseries 12} (2013) 036} [\href{https://arxiv.org/abs/1308.4754}{{\ttfamily 1308.4754}}].

\bibitem{Fujita:2014tja}
T.~Fujita, M.~Kawasaki and Y.~Tada, \emph{{Non-perturbative approach for curvature perturbations in stochastic $\delta N$ formalism}}, \href{https://doi.org/10.1088/1475-7516/2014/10/030}{\emph{JCAP} {\bfseries 10} (2014) 030} [\href{https://arxiv.org/abs/1405.2187}{{\ttfamily 1405.2187}}].

\bibitem{Vennin:2015hra}
V.~Vennin and A.A.~Starobinsky, \emph{{Correlation Functions in Stochastic Inflation}}, \href{https://doi.org/10.1140/epjc/s10052-015-3643-y}{\emph{Eur. Phys. J. C} {\bfseries 75} (2015) 413} [\href{https://arxiv.org/abs/1506.04732}{{\ttfamily 1506.04732}}].

\bibitem{Carr:1974nx}
B.J.~Carr and S.W.~Hawking, \emph{{Black holes in the early Universe}}, \href{https://doi.org/10.1093/mnras/168.2.399}{\emph{Mon. Not. Roy. Astron. Soc.} {\bfseries 168} (1974) 399}.

\bibitem{Carr:1975qj}
B.J.~Carr, \emph{{The Primordial black hole mass spectrum}}, \href{https://doi.org/10.1086/153853}{\emph{Astrophys. J.} {\bfseries 201} (1975) 1}.

\bibitem{Pattison:2017mbe}
C.~Pattison, V.~Vennin, H.~Assadullahi and D.~Wands, \emph{{Quantum diffusion during inflation and primordial black holes}}, \href{https://doi.org/10.1088/1475-7516/2017/10/046}{\emph{JCAP} {\bfseries 10} (2017) 046} [\href{https://arxiv.org/abs/1707.00537}{{\ttfamily 1707.00537}}].

\bibitem{Cruces:2018cvq}
D.~Cruces, C.~Germani and T.~Prokopec, \emph{{Failure of the stochastic approach to inflation beyond slow-roll}}, \href{https://doi.org/10.1088/1475-7516/2019/03/048}{\emph{JCAP} {\bfseries 03} (2019) 048} [\href{https://arxiv.org/abs/1807.09057}{{\ttfamily 1807.09057}}].

\bibitem{Ezquiaga:2018gbw}
J.M.~Ezquiaga and J.~Garc\'\i{}a-Bellido, \emph{{Quantum diffusion beyond slow-roll: implications for primordial black-hole production}}, \href{https://doi.org/10.1088/1475-7516/2018/08/018}{\emph{JCAP} {\bfseries 08} (2018) 018} [\href{https://arxiv.org/abs/1805.06731}{{\ttfamily 1805.06731}}].

\bibitem{Biagetti:2018pjj}
M.~Biagetti, G.~Franciolini, A.~Kehagias and A.~Riotto, \emph{{Primordial Black Holes from Inflation and Quantum Diffusion}}, \href{https://doi.org/10.1088/1475-7516/2018/07/032}{\emph{JCAP} {\bfseries 07} (2018) 032} [\href{https://arxiv.org/abs/1804.07124}{{\ttfamily 1804.07124}}].

\bibitem{Firouzjahi:2018vet}
H.~Firouzjahi, A.~Nassiri-Rad and M.~Noorbala, \emph{{Stochastic Ultra Slow Roll Inflation}}, \href{https://doi.org/10.1088/1475-7516/2019/01/040}{\emph{JCAP} {\bfseries 01} (2019) 040} [\href{https://arxiv.org/abs/1811.02175}{{\ttfamily 1811.02175}}].

\bibitem{Ezquiaga:2019ftu}
J.M.~Ezquiaga, J.~Garc\'\i{}a-Bellido and V.~Vennin, \emph{{The exponential tail of inflationary fluctuations: consequences for primordial black holes}}, \href{https://doi.org/10.1088/1475-7516/2020/03/029}{\emph{JCAP} {\bfseries 03} (2020) 029} [\href{https://arxiv.org/abs/1912.05399}{{\ttfamily 1912.05399}}].

\bibitem{Pattison:2019hef}
C.~Pattison, V.~Vennin, H.~Assadullahi and D.~Wands, \emph{{Stochastic inflation beyond slow roll}}, \href{https://doi.org/10.1088/1475-7516/2019/07/031}{\emph{JCAP} {\bfseries 07} (2019) 031} [\href{https://arxiv.org/abs/1905.06300}{{\ttfamily 1905.06300}}].

\bibitem{Prokopec:2019srf}
T.~Prokopec and G.~Rigopoulos, \emph{{\ensuremath{\Delta}N and the stochastic conveyor belt of ultra slow-roll inflation}}, \href{https://doi.org/10.1103/PhysRevD.104.083505}{\emph{Phys. Rev. D} {\bfseries 104} (2021) 083505} [\href{https://arxiv.org/abs/1910.08487}{{\ttfamily 1910.08487}}].

\bibitem{Ando:2020fjm}
K.~Ando and V.~Vennin, \emph{{Power spectrum in stochastic inflation}}, \href{https://doi.org/10.1088/1475-7516/2021/04/057}{\emph{JCAP} {\bfseries 04} (2021) 057} [\href{https://arxiv.org/abs/2012.02031}{{\ttfamily 2012.02031}}].

\bibitem{De:2020hdo}
A.~De and R.~Mahbub, \emph{{Numerically modeling stochastic inflation in slow-roll and beyond}}, \href{https://doi.org/10.1103/PhysRevD.102.123509}{\emph{Phys. Rev. D} {\bfseries 102} (2020) 123509} [\href{https://arxiv.org/abs/2010.12685}{{\ttfamily 2010.12685}}].

\bibitem{Figueroa:2020jkf}
D.G.~Figueroa, S.~Raatikainen, S.~Rasanen and E.~Tomberg, \emph{{Non-Gaussian Tail of the Curvature Perturbation in Stochastic Ultraslow-Roll Inflation: Implications for Primordial Black Hole Production}}, \href{https://doi.org/10.1103/PhysRevLett.127.101302}{\emph{Phys. Rev. Lett.} {\bfseries 127} (2021) 101302} [\href{https://arxiv.org/abs/2012.06551}{{\ttfamily 2012.06551}}].

\bibitem{Ballesteros:2020sre}
G.~Ballesteros, J.~Rey, M.~Taoso and A.~Urbano, \emph{{Stochastic inflationary dynamics beyond slow-roll and consequences for primordial black hole formation}}, \href{https://doi.org/10.1088/1475-7516/2020/08/043}{\emph{JCAP} {\bfseries 08} (2020) 043} [\href{https://arxiv.org/abs/2006.14597}{{\ttfamily 2006.14597}}].

\bibitem{Vennin:2020kng}
V.~Vennin, \emph{{Stochastic inflation and primordial black holes}}, Ph.D. thesis, U. Paris-Saclay, 6, 2020.
\newblock \href{https://arxiv.org/abs/2009.08715}{{\ttfamily 2009.08715}}.

\bibitem{Cruces:2021iwq}
D.~Cruces and C.~Germani, \emph{{Stochastic inflation at all order in slow-roll parameters: Foundations}}, \href{https://doi.org/10.1103/PhysRevD.105.023533}{\emph{Phys. Rev. D} {\bfseries 105} (2022) 023533} [\href{https://arxiv.org/abs/2107.12735}{{\ttfamily 2107.12735}}].

\bibitem{Rigopoulos:2021nhv}
G.~Rigopoulos and A.~Wilkins, \emph{{Inflation is always semi-classical: diffusion domination overproduces Primordial Black Holes}}, \href{https://doi.org/10.1088/1475-7516/2021/12/027}{\emph{JCAP} {\bfseries 12} (2021) 027} [\href{https://arxiv.org/abs/2107.05317}{{\ttfamily 2107.05317}}].

\bibitem{Pattison:2021oen}
C.~Pattison, V.~Vennin, D.~Wands and H.~Assadullahi, \emph{{Ultra-slow-roll inflation with quantum diffusion}}, \href{https://doi.org/10.1088/1475-7516/2021/04/080}{\emph{JCAP} {\bfseries 04} (2021) 080} [\href{https://arxiv.org/abs/2101.05741}{{\ttfamily 2101.05741}}].

\bibitem{Achucarro:2021pdh}
A.~Achucarro, S.~Cespedes, A.-C.~Davis and G.A.~Palma, \emph{{The hand-made tail: non-perturbative tails from multifield inflation}}, \href{https://doi.org/10.1007/JHEP05(2022)052}{\emph{JHEP} {\bfseries 05} (2022) 052} [\href{https://arxiv.org/abs/2112.14712}{{\ttfamily 2112.14712}}].

\bibitem{Hooshangi:2021ubn}
S.~Hooshangi, M.H.~Namjoo and M.~Noorbala, \emph{{Rare events are nonperturbative: Primordial black holes from heavy-tailed distributions}}, \href{https://doi.org/10.1016/j.physletb.2022.137400}{\emph{Phys. Lett. B} {\bfseries 834} (2022) 137400} [\href{https://arxiv.org/abs/2112.04520}{{\ttfamily 2112.04520}}].

\bibitem{Tomberg:2021xxv}
E.~Tomberg, \emph{{A numerical approach to stochastic inflation and primordial black holes}}, \href{https://doi.org/10.1088/1742-6596/2156/1/012010}{\emph{J. Phys. Conf. Ser.} {\bfseries 2156} (2021) 012010} [\href{https://arxiv.org/abs/2110.10684}{{\ttfamily 2110.10684}}].

\bibitem{Figueroa:2021zah}
D.G.~Figueroa, S.~Raatikainen, S.~Rasanen and E.~Tomberg, \emph{{Implications of stochastic effects for primordial black hole production in ultra-slow-roll inflation}}, \href{https://doi.org/10.1088/1475-7516/2022/05/027}{\emph{JCAP} {\bfseries 05} (2022) 027} [\href{https://arxiv.org/abs/2111.07437}{{\ttfamily 2111.07437}}].

\bibitem{Tada:2021zzj}
Y.~Tada and V.~Vennin, \emph{{Statistics of coarse-grained cosmological fields in stochastic inflation}}, \href{https://doi.org/10.1088/1475-7516/2022/02/021}{\emph{JCAP} {\bfseries 02} (2022) 021} [\href{https://arxiv.org/abs/2111.15280}{{\ttfamily 2111.15280}}].

\bibitem{Cruces:2022imf}
D.~Cruces, \emph{{Review on Stochastic Approach to Inflation}}, \href{https://doi.org/10.3390/universe8060334}{\emph{Universe} {\bfseries 8} (2022) 334} [\href{https://arxiv.org/abs/2203.13852}{{\ttfamily 2203.13852}}].

\bibitem{Ahmadi:2022lsm}
N.~Ahmadi, M.~Noorbala, N.~Feyzabadi, F.~Eghbalpoor and Z.~Ahmadi, \emph{{Quantum diffusion in sharp transition to non-slow-roll phase}}, \href{https://doi.org/10.1088/1475-7516/2022/08/078}{\emph{JCAP} {\bfseries 08} (2022) 078} [\href{https://arxiv.org/abs/2207.10578}{{\ttfamily 2207.10578}}].

\bibitem{Animali:2022otk}
C.~Animali and V.~Vennin, \emph{{Primordial black holes from stochastic tunnelling}}, \href{https://doi.org/10.1088/1475-7516/2023/02/043}{\emph{JCAP} {\bfseries 02} (2023) 043} [\href{https://arxiv.org/abs/2210.03812}{{\ttfamily 2210.03812}}].

\bibitem{Jackson:2022unc}
J.H.P.~Jackson, H.~Assadullahi, K.~Koyama, V.~Vennin and D.~Wands, \emph{{Numerical simulations of stochastic inflation using importance sampling}}, \href{https://doi.org/10.1088/1475-7516/2022/10/067}{\emph{JCAP} {\bfseries 10} (2022) 067} [\href{https://arxiv.org/abs/2206.11234}{{\ttfamily 2206.11234}}].

\bibitem{Tomberg:2022mkt}
E.~Tomberg, \emph{{Numerical stochastic inflation constrained by frozen noise}}, \href{https://doi.org/10.1088/1475-7516/2023/04/042}{\emph{JCAP} {\bfseries 04} (2023) 042} [\href{https://arxiv.org/abs/2210.17441}{{\ttfamily 2210.17441}}].

\bibitem{Tomberg:2023kli}
E.~Tomberg, \emph{{Stochastic constant-roll inflation and primordial black holes}}, \href{https://doi.org/10.1103/PhysRevD.108.043502}{\emph{Phys. Rev. D} {\bfseries 108} (2023) 043502} [\href{https://arxiv.org/abs/2304.10903}{{\ttfamily 2304.10903}}].

\bibitem{Asadi:2023flu}
K.~Asadi, A.~Nassiri-Rad and H.~Firouzjahi, \emph{{Stochastic multiple fields inflation: Diffusion dominated regime}}, \href{https://doi.org/10.1103/PhysRevD.108.123537}{\emph{Phys. Rev. D} {\bfseries 108} (2023) 123537} [\href{https://arxiv.org/abs/2304.00577}{{\ttfamily 2304.00577}}].

\bibitem{Briaud:2023eae}
V.~Briaud and V.~Vennin, \emph{{Uphill inflation}}, \href{https://doi.org/10.1088/1475-7516/2023/06/029}{\emph{JCAP} {\bfseries 06} (2023) 029} [\href{https://arxiv.org/abs/2301.09336}{{\ttfamily 2301.09336}}].

\bibitem{Mishra:2023lhe}
S.S.~Mishra, E.J.~Copeland and A.M.~Green, \emph{{Primordial black holes and stochastic inflation beyond slow roll. Part I. Noise matrix elements}}, \href{https://doi.org/10.1088/1475-7516/2023/09/005}{\emph{JCAP} {\bfseries 09} (2023) 005} [\href{https://arxiv.org/abs/2303.17375}{{\ttfamily 2303.17375}}].

\bibitem{Raatikainen:2023bzk}
S.~Raatikainen, S.~R\"as\"anen and E.~Tomberg, \emph{{Primordial Black Hole Compaction Function from Stochastic Fluctuations in Ultraslow-Roll Inflation}}, \href{https://doi.org/10.1103/PhysRevLett.133.121403}{\emph{Phys. Rev. Lett.} {\bfseries 133} (2024) 121403} [\href{https://arxiv.org/abs/2312.12911}{{\ttfamily 2312.12911}}].

\bibitem{Vennin:2024yzl}
V.~Vennin and D.~Wands, \emph{{Quantum diffusion and large primordial perturbations from inflation}},  \href{https://arxiv.org/abs/2402.12672}{{\ttfamily 2402.12672}}.

\bibitem{Jackson:2024aoo}
J.H.P.~Jackson, H.~Assadullahi, A.D.~Gow, K.~Koyama, V.~Vennin and D.~Wands, \emph{{Stochastic inflation beyond slow roll: noise modelling and importance sampling}},  \href{https://arxiv.org/abs/2410.13683}{{\ttfamily 2410.13683}}.

\bibitem{Ito:1944}
K.~It{\^o}, \emph{Stochastic integral}, \href{https://doi.org/10.3792/pia/1195572786}{\emph{Proceedings of the Imperial Academy} {\bfseries 20} (1944) 519}.

\bibitem{Stratonovich:1966}
R.L.~Stratonovich, \emph{A new representation for stochastic integrals and equations}, \href{https://doi.org/10.1137/0304028}{\emph{SIAM Journal on Control} {\bfseries 4} (1966) 362} [\href{https://arxiv.org/abs/https://doi.org/10.1137/0304028}{{\ttfamily https://doi.org/10.1137/0304028}}].

\bibitem{ItoStratoWebpage}
``It{\^o} and {{Stratonovich}}; a guide for the perplexed - {{OATML}}.'' https://oatml.cs.ox.ac.uk/blog/2022/03/22/ito-strat.html.

\bibitem{Kampen:1981}
N.G.~{van Kampen}, \emph{It{\^o} versus {{Stratonovich}}}, \href{https://doi.org/10.1007/BF01007642}{\emph{Journal of Statistical Physics} {\bfseries 24} (1981) 175}.

\bibitem{Mannella:2012}
R.~Mannella and P.V.E.~McCLINTOCK, \emph{{{IT{\^O} VERSUS STRATONOVICH}}: 30 {{YEARS LATER}}}, \href{https://doi.org/10.1142/S021947751240010X}{\emph{Fluctuation and Noise Letters} {\bfseries 11} (2012) 1240010}.

\bibitem{Salopek:1990re}
D.S.~Salopek and J.R.~Bond, \emph{{Stochastic inflation and nonlinear gravity}}, \href{https://doi.org/10.1103/PhysRevD.43.1005}{\emph{Phys. Rev. D} {\bfseries 43} (1991) 1005}.

\bibitem{Mezhlumian:1991hw}
A.~Mezhlumian and A.A.~Starobinsky, \emph{{Stochastic inflation: New results}},  in \emph{{The First International A.D. Sakharov Conference on Physics}}, 10, 1991 [\href{https://arxiv.org/abs/astro-ph/9406045}{{\ttfamily astro-ph/9406045}}].

\bibitem{Linde:1993xx}
A.D.~Linde, D.A.~Linde and A.~Mezhlumian, \emph{{From the Big Bang theory to the theory of a stationary universe}}, \href{https://doi.org/10.1103/PhysRevD.49.1783}{\emph{Phys. Rev. D} {\bfseries 49} (1994) 1783} [\href{https://arxiv.org/abs/gr-qc/9306035}{{\ttfamily gr-qc/9306035}}].

\bibitem{Winitzki:1995pg}
S.~Winitzki and A.~Vilenkin, \emph{{Uncertainties of predictions in models of eternal inflation}}, \href{https://doi.org/10.1103/PhysRevD.53.4298}{\emph{Phys. Rev. D} {\bfseries 53} (1996) 4298} [\href{https://arxiv.org/abs/gr-qc/9510054}{{\ttfamily gr-qc/9510054}}].

\bibitem{Garriga:1997ef}
J.~Garriga and A.~Vilenkin, \emph{{Recycling universe}}, \href{https://doi.org/10.1103/PhysRevD.57.2230}{\emph{Phys. Rev. D} {\bfseries 57} (1998) 2230} [\href{https://arxiv.org/abs/astro-ph/9707292}{{\ttfamily astro-ph/9707292}}].

\bibitem{Vilenkin:1999kd}
A.~Vilenkin, \emph{{On the factor ordering problem in stochastic inflation}}, \href{https://doi.org/10.1103/PhysRevD.59.123506}{\emph{Phys. Rev. D} {\bfseries 59} (1999) 123506} [\href{https://arxiv.org/abs/gr-qc/9902007}{{\ttfamily gr-qc/9902007}}].

\bibitem{Winitzki:2008zz}
S.~Winitzki, \emph{{Eternal inflation}} (2008), \href{https://doi.org/10.1142/6923}{10.1142/6923}.

\bibitem{Tokuda:2017fdh}
J.~Tokuda and T.~Tanaka, \emph{{Statistical nature of infrared dynamics on de Sitter background}}, \href{https://doi.org/10.1088/1475-7516/2018/02/014}{\emph{JCAP} {\bfseries 02} (2018) 014} [\href{https://arxiv.org/abs/1708.01734}{{\ttfamily 1708.01734}}].

\bibitem{Pinol:2018euk}
L.~Pinol, S.~Renaux-Petel and Y.~Tada, \emph{{Inflationary stochastic anomalies}}, \href{https://doi.org/10.1088/1361-6382/ab097f}{\emph{Class. Quant. Grav.} {\bfseries 36} (2019) 07LT01} [\href{https://arxiv.org/abs/1806.10126}{{\ttfamily 1806.10126}}].

\bibitem{Pinol:2020cdp}
L.~Pinol, S.~Renaux-Petel and Y.~Tada, \emph{{A manifestly covariant theory of multifield stochastic inflation in phase space: solving the discretisation ambiguity in stochastic inflation}}, \href{https://doi.org/10.1088/1475-7516/2021/04/048}{\emph{JCAP} {\bfseries 04} (2021) 048} [\href{https://arxiv.org/abs/2008.07497}{{\ttfamily 2008.07497}}].

\bibitem{Malik:2008im}
K.A.~Malik and D.~Wands, \emph{{Cosmological perturbations}}, \href{https://doi.org/10.1016/j.physrep.2009.03.001}{\emph{Phys. Rept.} {\bfseries 475} (2009) 1} [\href{https://arxiv.org/abs/0809.4944}{{\ttfamily 0809.4944}}].

\bibitem{Birrell:1982ix}
N.D.~Birrell and P.C.W.~Davies, \emph{{Quantum Fields in Curved Space}}, Cambridge Monographs on Mathematical Physics, Cambridge University Press, Cambridge, UK (1982), \href{https://doi.org/10.1017/CBO9780511622632}{10.1017/CBO9780511622632}.

\bibitem{Grain:2019vnq}
J.~Grain and V.~Vennin, \emph{{Canonical transformations and squeezing formalism in cosmology}}, \href{https://doi.org/10.1088/1475-7516/2020/02/022}{\emph{JCAP} {\bfseries 02} (2020) 022} [\href{https://arxiv.org/abs/1910.01916}{{\ttfamily 1910.01916}}].

\bibitem{Kloeden:1992}
P.E.~Kloeden and E.~Platen, \emph{Numerical Solution of Stochastic Differential Equations}, Springer Berlin Heidelberg, Berlin, Heidelberg (1992), \href{https://doi.org/10.1007/978-3-662-12616-5}{10.1007/978-3-662-12616-5}.

\bibitem{Foster:2020}
J.M.~Foster, \emph{Numerical approximations for stochastic differential equations}, Ph.D. thesis, University of Oxford, 2020.

\bibitem{Goncharov:1987ir}
A.S.~Goncharov, A.D.~Linde and V.F.~Mukhanov, \emph{{The Global Structure of the Inflationary Universe}}, \href{https://doi.org/10.1142/S0217751X87000211}{\emph{Int. J. Mod. Phys. A} {\bfseries 2} (1987) 561}.

\bibitem{Guth:2000ka}
A.H.~Guth, \emph{{Inflation and eternal inflation}}, \href{https://doi.org/10.1016/S0370-1573(00)00037-5}{\emph{Phys. Rept.} {\bfseries 333} (2000) 555} [\href{https://arxiv.org/abs/astro-ph/0002156}{{\ttfamily astro-ph/0002156}}].

\bibitem{Linde:2015edk}
A.~Linde, \emph{{A brief history of the multiverse}}, \href{https://doi.org/10.1088/1361-6633/aa50e4}{\emph{Rept. Prog. Phys.} {\bfseries 80} (2017) 022001} [\href{https://arxiv.org/abs/1512.01203}{{\ttfamily 1512.01203}}].

\bibitem{Linde:1986fc}
A.D.~Linde, \emph{{ETERNAL CHAOTIC INFLATION}}, \href{https://doi.org/10.1142/S0217732386000129}{\emph{Mod. Phys. Lett. A} {\bfseries 1} (1986) 81}.

\bibitem{Dimopoulos:2017ged}
K.~Dimopoulos, \emph{{Ultra slow-roll inflation demystified}}, \href{https://doi.org/10.1016/j.physletb.2017.10.066}{\emph{Phys. Lett. B} {\bfseries 775} (2017) 262} [\href{https://arxiv.org/abs/1707.05644}{{\ttfamily 1707.05644}}].

\bibitem{Lyth:2009zz}
D.H.~Lyth and A.R.~Liddle, \emph{{The primordial density perturbation: Cosmology, inflation and the origin of structure}} (2009).

\bibitem{Karam:2022nym}
A.~Karam, N.~Koivunen, E.~Tomberg, V.~Vaskonen and H.~Veerm\"ae, \emph{{Anatomy of single-field inflationary models for primordial black holes}}, \href{https://doi.org/10.1088/1475-7516/2023/03/013}{\emph{JCAP} {\bfseries 03} (2023) 013} [\href{https://arxiv.org/abs/2205.13540}{{\ttfamily 2205.13540}}].

\bibitem{Figueroa:2020rrl}
D.G.~Figueroa, A.~Florio, F.~Torrenti and W.~Valkenburg, \emph{{The art of simulating the early Universe -- Part I}}, \href{https://doi.org/10.1088/1475-7516/2021/04/035}{\emph{JCAP} {\bfseries 04} (2021) 035} [\href{https://arxiv.org/abs/2006.15122}{{\ttfamily 2006.15122}}].

\bibitem{Cohen:2021jbo}
T.~Cohen, D.~Green and A.~Premkumar, \emph{{A tail of eternal inflation}}, \href{https://doi.org/10.21468/SciPostPhys.14.5.109}{\emph{SciPost Phys.} {\bfseries 14} (2023) 109} [\href{https://arxiv.org/abs/2111.09332}{{\ttfamily 2111.09332}}].

\bibitem{Cohen:2022clv}
T.~Cohen, D.~Green and A.~Premkumar, \emph{{Large deviations in the early Universe}}, \href{https://doi.org/10.1103/PhysRevD.107.083501}{\emph{Phys. Rev. D} {\bfseries 107} (2023) 083501} [\href{https://arxiv.org/abs/2212.02535}{{\ttfamily 2212.02535}}].

\bibitem{Jackson:2023obv}
J.H.P.~Jackson, H.~Assadullahi, A.D.~Gow, K.~Koyama, V.~Vennin and D.~Wands, \emph{{The separate-universe approach and sudden transitions during inflation}}, \href{https://doi.org/10.1088/1475-7516/2024/05/053}{\emph{JCAP} {\bfseries 05} (2024) 053} [\href{https://arxiv.org/abs/2311.03281}{{\ttfamily 2311.03281}}].

\bibitem{Zakamska:2015nys}
N.L.~Zakamska, \emph{{Theory of Special Relativity}},  \href{https://arxiv.org/abs/1511.02121}{{\ttfamily 1511.02121}}.

\bibitem{Peskin:1995ev}
M.E.~Peskin and D.V.~Schroeder, \emph{{An Introduction to quantum field theory}}, Addison-Wesley, Reading, USA (1995), \href{https://doi.org/10.1201/9780429503559}{10.1201/9780429503559}.

\bibitem{Cai:2018dkf}
Y.-F.~Cai, X.~Chen, M.H.~Namjoo, M.~Sasaki, D.-G.~Wang and Z.~Wang, \emph{{Revisiting non-Gaussianity from non-attractor inflation models}}, \href{https://doi.org/10.1088/1475-7516/2018/05/012}{\emph{JCAP} {\bfseries 05} (2018) 012} [\href{https://arxiv.org/abs/1712.09998}{{\ttfamily 1712.09998}}].

\bibitem{Atal:2019cdz}
V.~Atal, J.~Garriga and A.~Marcos-Caballero, \emph{{Primordial black hole formation with non-Gaussian curvature perturbations}}, \href{https://doi.org/10.1088/1475-7516/2019/09/073}{\emph{JCAP} {\bfseries 09} (2019) 073} [\href{https://arxiv.org/abs/1905.13202}{{\ttfamily 1905.13202}}].

\bibitem{Atal:2019erb}
V.~Atal, J.~Cid, A.~Escriv\`a and J.~Garriga, \emph{{PBH in single field inflation: the effect of shape dispersion and non-Gaussianities}}, \href{https://doi.org/10.1088/1475-7516/2020/05/022}{\emph{JCAP} {\bfseries 05} (2020) 022} [\href{https://arxiv.org/abs/1908.11357}{{\ttfamily 1908.11357}}].

\bibitem{Biagetti:2021eep}
M.~Biagetti, V.~De~Luca, G.~Franciolini, A.~Kehagias and A.~Riotto, \emph{{The formation probability of primordial black holes}}, \href{https://doi.org/10.1016/j.physletb.2021.136602}{\emph{Phys. Lett. B} {\bfseries 820} (2021) 136602} [\href{https://arxiv.org/abs/2105.07810}{{\ttfamily 2105.07810}}].

\bibitem{Hooshangi:2022lao}
S.~Hooshangi, A.~Talebian, M.H.~Namjoo and H.~Firouzjahi, \emph{{Multiple field ultraslow-roll inflation: Primordial black holes from straight bulk and distorted boundary}}, \href{https://doi.org/10.1103/PhysRevD.105.083525}{\emph{Phys. Rev. D} {\bfseries 105} (2022) 083525} [\href{https://arxiv.org/abs/2201.07258}{{\ttfamily 2201.07258}}].

\bibitem{Pi:2022ysn}
S.~Pi and M.~Sasaki, \emph{{Logarithmic Duality of the Curvature Perturbation}}, \href{https://doi.org/10.1103/PhysRevLett.131.011002}{\emph{Phys. Rev. Lett.} {\bfseries 131} (2023) 011002} [\href{https://arxiv.org/abs/2211.13932}{{\ttfamily 2211.13932}}].

\bibitem{Firouzjahi:2023ahg}
H.~Firouzjahi and A.~Riotto, \emph{{Primordial Black Holes and loops in single-field inflation}}, \href{https://doi.org/10.1088/1475-7516/2024/02/021}{\emph{JCAP} {\bfseries 02} (2024) 021} [\href{https://arxiv.org/abs/2304.07801}{{\ttfamily 2304.07801}}].

\bibitem{Hooshangi:2023kss}
S.~Hooshangi, M.H.~Namjoo and M.~Noorbala, \emph{{Tail diversity from inflation}}, \href{https://doi.org/10.1088/1475-7516/2023/09/023}{\emph{JCAP} {\bfseries 09} (2023) 023} [\href{https://arxiv.org/abs/2305.19257}{{\ttfamily 2305.19257}}].

\bibitem{Pi:2024jwt}
S.~Pi, \emph{{Non-Gaussianities in primordial black hole formation and induced gravitational waves}},  \href{https://arxiv.org/abs/2404.06151}{{\ttfamily 2404.06151}}.

\bibitem{Artigas:2024xhc}
D.~Artigas, S.~Pi and T.~Tanaka, \emph{{Extended $\delta N$ formalism}},  \href{https://arxiv.org/abs/2408.09964}{{\ttfamily 2408.09964}}.

\bibitem{Inui:2024sce}
R.~Inui, H.~Motohashi, S.~Pi, Y.~Tada and S.~Yokoyama, \emph{{Constant roll and non-Gaussian tail in light of logarithmic duality}}, \href{https://doi.org/10.1088/1475-7516/2025/02/042}{\emph{JCAP} {\bfseries 02} (2025) 042} [\href{https://arxiv.org/abs/2409.13500}{{\ttfamily 2409.13500}}].

\bibitem{Ballesteros:2024pwn}
G.~Ballesteros, T.~Konstandin, A.~P\'erez~Rodr\'\i{}guez, M.~Pierre and J.~Rey, \emph{{Non-Gaussian tails without stochastic inflation}}, \href{https://doi.org/10.1088/1475-7516/2024/11/013}{\emph{JCAP} {\bfseries 11} (2024) 013} [\href{https://arxiv.org/abs/2406.02417}{{\ttfamily 2406.02417}}].

\bibitem{Tokuda:2018eqs}
J.~Tokuda and T.~Tanaka, \emph{{Can all the infrared secular growth really be understood as increase of classical statistical variance?}}, \href{https://doi.org/10.1088/1475-7516/2018/11/022}{\emph{JCAP} {\bfseries 11} (2018) 022} [\href{https://arxiv.org/abs/1806.03262}{{\ttfamily 1806.03262}}].

\bibitem{Launay:2024qsm}
Y.L.~Launay, G.I.~Rigopoulos and E.P.S.~Shellard, \emph{{Stochastic inflation in general relativity}}, \href{https://doi.org/10.1103/PhysRevD.109.123523}{\emph{Phys. Rev. D} {\bfseries 109} (2024) 123523} [\href{https://arxiv.org/abs/2401.08530}{{\ttfamily 2401.08530}}].

\bibitem{Martingales:2023}
{\'E}.~Rold{\'a}n, I.~Neri, R.~Chetrite, S.~Gupta, S.~Pigolotti, F.~J{\"u}licher et~al., \emph{Martingales for physicists: a treatise on stochastic thermodynamics and beyond}, \href{https://doi.org/10.1080/00018732.2024.2317494}{\emph{Advances in Physics} {\bfseries 72} (2023) 1} [\href{https://arxiv.org/abs/https://doi.org/10.1080/00018732.2024.2317494}{{\ttfamily https://doi.org/10.1080/00018732.2024.2317494}}].

\end{thebibliography}\endgroup

\end{document}